\documentclass{article}
\usepackage{spconf,amsmath,graphicx}
\usepackage{amsfonts}

\usepackage{tikz}  
\usepackage{adjustbox}
\usepackage[section]{placeins} 

\title{Distribution Conditional Denoising: A Flexible Discriminative Image Denoiser}
\name{Anthony Kelly}
\address{University of Limerick, Limerick, Ireland}
\begin{document}
%
\maketitle
\begin{abstract}
A flexible discriminative image denoiser is introduced in which multi-task learning methods are applied to a densoising FCN based on U-Net. The activations of the U-Net model are modified by affine transforms that are a learned function of  conditioning inputs. The learning procedure for multiple noise types and levels involves applying a distribution of noise parameters during training to the conditioning inputs, with the same noise parameters applied to a noise generating layer at the input (similar to the approach taken in a denoising autoencoder). It is shown that this flexible denoising model achieves state of the art performance on images corrupted with Gaussian and Poisson noise. It has also been shown that this conditional training method can generalise a fixed noise level U-Net denoiser to a variety of noise levels.
\end{abstract}

\section{Introduction}
\label{sec:intro}
The process of capturing and processing images involves inevitable contamination by noise. From the Poisson nature of photon detection by a sensor to Gaussian electronic thermal noise, the noise contamination affects visual perception of an image. That perception may be aesthetic in camera images, but in medical imaging noise can affect the ability of a clinician to visibly identify a potentially harmful lesion in an X-ray or CT-scan. Improving image quality can improve diagnosis with lower levels of radiation exposure \cite{Huda2015}.

In general, effective denoising methods should achieve smoothing in flat areas but edges and textures should be preserved and artifacts should not be added to the image. Many methods assume a Gaussian Noise model and perform best when provided with an accurate estimate of the noise level, requiring  prior knowledge of the noise level of the input image. There is a clear deterioration in performance when the noise level estimate provided to the denoiser is fixed so that it does not match the noise level of the input image as the noise level is varied.  If the noise level provided to the denoiser is over estimated versus the image then the image will be oversmoothed, resulting in poor edge rendering and artifacts \cite{Burger2012}.  For discriminative methods, matching the model to the input noise requires many models, one for each level of input noise and each noise type, with poor generalisation evident for noise distributions not used in training \cite{Plotz2017}. Generally discriminative methods have assumed only Gaussian noise.

 For the purposes of flexibility and computational efficiency it is desirable that a single model can  operate effectively on different levels of input noise and different noise types. In medical imaging for example, the level of noise depends on the machine parameters and the dose of radiation involved in capturing the image \cite{Leuschner2019} and low-dose X-ray can result in higher relative noise that makes diagnosis of lesions difficult \cite{Lee2018,Huda2015}. Furthermore, the noise generating mechanism in medical images are typically characterised by a Poisson-Gaussian model where the parameters of the noise generation model vary according to the type of equipment involved \cite{Lee2018}, and may  not be accurately known. Employing a single denoise model for many practical noise levels and types is a clear advantage in this and many other scenarios. A fast execution time, and variable noise level capability with a low memory footprint are desirable aspects of a denoising method. An extension of neural networks denoisers to multiple noise levels and types is advantageous since their speed of execution at inference time is generally faster than collaborative filtering based methods such BM3D \cite{Dabov2007a}.
 
In this paper we introduce a method to generalise the denoising performance of a denoising Neural Network to encompass a range of noise parameters. In particular, multi-task learning methods are applied to a densoising FCN based on U-Net \cite{Ronneberger2015} so that the denoising model is conditioned on a distribution of parameters of a Poisson-Gaussian noise model. In this way, the denoising U-Net model is generalised to be a function of the conditioning inputs parameterising the Poisson-Gaussian noise model.

\section{Related Work}
\label{sec:related}
Since the noise free image falling on a sensor cannot be observed directly, image denoising is defined as an inverse problem in which the noise corrupted image must be mapped back to the noise free image. The goal of the solution to the inverse problem in image processing is to find the latent image from the observation, which is generally a sparse and noisy observation. Being sparse and noisy, this is an ill posed problem \cite{Sun2018} and the solution involves optimization of a regularised minimisation problem \cite{Heide2014}. Denoising often introduces artifacts or removes the fine details of the image \cite{Buades2005,Chen2017b}.

Spatial domain filtering methods typically employ filters to smooth the image, but can suffer from edge blurring \cite{Tomasi1998}. Variational spatial domain methods introduce a prior by solving for a MAP estimate of the image in order to produce more natural images with less over smoothing, but optimisation can be computationally expensive and time consuming. Grouping and collaborative filtering is applied by BM3D   \cite{Dabov2007} and is considered the benchmark to which other methods are compared \cite{Chen2018,Heide2014,Remez2017a}. Although many methods have claimed superior results to BM3D, it has been found that BM3D (and CBM3D, its colour image version), outperforms many of those methods on natural images \cite{Plotz2017}.
Although a Gaussian noise model is widely used for modelling noise in images, a Poisson-Gaussian model is a more appropriate noise model in many cases, including medical imaging \cite{Anaya2018,Foi2008}. It has been shown that BM3D does not generalise well to Poisson-Gaussian noise \cite{Thanh2019}.
 
A typical inverse problem is shown in equation (\ref{equ:im_obs_mdl}), involving an unknown latent image denoted by the vector x, vector z denotes the sensor image, matrix A is a transformation matrix and $\eta$ represents noise. The observation model is:

\begin{equation}\label{equ:im_obs_mdl}
    z=Ax+\eta
\end{equation}

With the optimisation problem consisting of a fidelity operator and a regularization operator as follows:

\begin{equation}\label{soln_im_obs_mdl}
    min_x\frac{1}{2}||z-Ax||^2_2 + \Gamma(x) 
\end{equation}

Such optimisation problems have been solved by the application of Half Quadratic Splitting \cite{Zoran2011}, allowing the fidelity operator and regularization operator to be separately solved. Proximal operators may also be employed \cite{Parikh2013} as demonstrated in FlexISP \cite{Heide2014} where it is shown that any Gaussian denoiser may be expressed as a proximal operator, including state-of the art methods such as BM3D. The solution of the fidelity operator may be solved as a least squares problem via the Conjugate Gradient algorithm \cite{Heide2014}. The advantage of the Proximal optimization in FlexISP is the ability to define different elements of the ISP pipeline via the transformation matrix A. Denoising is achieved when A is the Identity matrix. Deblurring can be achieved when A is a Gaussian blur. Incorporating downsampling and deblur in A yields Super Resolution. A disadvantage of FlexISP is the computation time and the reliance on a regularization prior.  Demosaicing takes between 312 seconds and 6 seconds depending on processing performance (CPU, GPU) and complexity of the denoiser. The choice of denoising prior affects the quality of the latent image, introducing artefacts.

\subsection{CNN Approaches}
Convolutional Neural Networks have been applied to the inverse problem \cite{Zhang2017a} with the CNN learning the optimal denoising prior whilst the fidelity operator is solved via HQS \cite{Zoran2011}. Bespoke CNN structures have been proposed for the end to end ISP where low level features and high level features are extracted via different paths before being combined \cite{Schwartz2018}.

CNNs have been applied where the CNN must learn the solution to the inverse problem without the use of priors, to jointly solve a complete end to end pipeline involving colour transform, demosaicing, denoising and image enhancement \cite{Chen2018}, producing good results on extremely dark images which traditionally produce very poor results in an ISP. In particular, fully convolutional networks (FCN) have been employed \cite{Shelhamer2017} and various architectures such as U-Net \cite{Ronneberger2015} and CAN \cite{Chen2017} have been compared, with U-Net providing better results. Dense neural networks have also been applied to the problem producing promising results compared to BM3D on realistic noise types but with very extensive computation times \cite{Burger2012}.

Prior knowledge of the noise level of the input image is an important factor in the denoising performance of discriminative denoising methods \cite{Burger2012}, where it is common for multiple models to be trained under various levels of input noise. Poor generalisation is evident for noise distributions not used in training \cite{Plotz2017}.

The problem of developing a single model discriminative denoiser has been addressed by adding the noise level as an additional input to the neural network to accommodate various noise levels  \cite{Burger2012}, but this was largely unsuccessful. Similarly, \cite{Zhang2018} added  a noise level map as a feature to a CNN based denoiser, where the noise map allows for noise variations in and across images. This network proved to be difficult to train with respect to the noise map. Several residual denoising discriminator methods have been proposed \cite{Zhang2017a} some of which require prior estimates of the noise level (DnCNN-S), whilst others (DnCNN-B and CDnCNN-B ), allow blind denoising, where prior noise estimates are not required or given, training instead on a distribution of noisy images. However, these models are applicable only to Gaussian noise, do not generalise well to real world noisy images and are limited in range of the noise standard deviation ([0,55] where $X\in[0,255]$) \cite{Zhang2018}. FFDNet \cite{Zhang2018} is a discriminative denoising method that attempts to address limitations of previous methods involving flexibility to deal with various noise levels in the same model or across the image, speed of execution and memory requirements. The flexibility was introduced by the addition of a Noise Map (M) feature to the training data. In comparisons, FFDNet was found to marginally outperform BM3D, and overall FFDNet was considered to produce "the best perceptual quality of denoised images" \cite{Fan2019}, assuming an AWGN noise model. Although FFDNET slightly out performed other methods based on PSNR comparisons, the inclusion of the the Noise Map M, made the model difficult to train and failed to control the trade-off between denoising and oversmoothing that was anticipated \cite{Zhang2018}. Despite the Noise Map allowing various noise levels, white Gaussian noise is assumed. Training time was reported to be 2 days on an Nvidia Titan X Pascal GPU.

There is some evidence that the structure of the CNN itself imposes an implicit prior on the image restoration and denoising problem. It has been shown that the learning curves of a U-Net FCN converged faster for natural images, implying that the structure of the FCN suits natural image restoration \cite{Lempitsky2018}. Limiting iterations of training helped the network reconstruct natural images better, indicating that high levels of detail, including noise, are learned later in the learning process. This means a previously untrained network can learn a standard inverse image problem (e.g. denoising, super-resolution, inpainting), showing that the network structure serves as a prior that does not rely on learning from a training dataset.

Although performance on image processing metrics is often considered to be the main differentiating factor in architecture selection, there are other important factors that favour the use of CNNs: i) CNNs provide a reusable architecture decoupled from the problem; ii) scale as compute power grows; iii) fast operation at inference time \cite{McCann2017}. Therefore, a CNN based denoising architecture that can be trained to denoise images corrupted with a range of various noise parameter values and not limited to Gaussian noise is of value.

\subsection{Multi-task Learning}
The literature of Multi-task learning is concerned with generalising the performance of a model to make it applicable to a wider, more general class of problems \cite{Ruder}. In solving multi-task problems, Conditional Normalization has been utilized as an approach that adapts a model in response to another input. For example, by scaling and shifting the parameters involved in layer normalization, an acoustic model may be adapted to different speakers and environments \cite{Kim}. The approach taken in the method of this paper is to generalise the performance of a discriminative denoiser to the general problem of denoising over a distribution of parameters of a more general noise model such as a Poisson-Gaussian model. 
 
 \subsection{Dynamic Normalisation}
Layer Normalisation in Deep Neural networks is concerned with reducing covariance shifts by learning the mean and variance of each layer's activations $a\in\mathbb{R}^{N}$ \cite{Ba2016}. It can be defined as a linear mapping function as follows:
\begin{equation}\label{LN_LayerNorm}
    LN(a;\alpha,\beta)=\alpha\odot\left(\frac{a-\mu}{\sigma}\right)+\beta
\end{equation}
\begin{equation*}
    \mu =\frac{1}{N}\sum_{i=1}^{N}a_i
\end{equation*}
\begin{equation*}
    \sigma=\sqrt{\frac{1}{N}\sum_{i=1}^{N}\left(a_i-\mu_i\right)^2}
\end{equation*}
where $a_i$ represents elements of the activations a, with $\mu$ and $\sigma$ representing the mean and standard deviation across the elements of a. The symbol $\odot$ represents the Hadamard or element wise product. The parameters $\alpha$ and $\beta$ are learned parameters that scale and shift the activity values respectively. When $\alpha=\sigma$ and $\beta=\mu$ the activations will be unscaled and unshifted, for example. Other values scale and shift the activations by various amounts. An advantage of Layer Normalisation over Batch Normalization \cite{Ioffe2015} is that it can be applied in RNNs and is operational in both training and inference \cite{Ba2016}.

Layer Normalisation may form the basis for an adaptive model whereby the scaling and shifting parameters are adjusted based on input data $x_i$ , rather than the layer activations:
\begin{equation}
    \alpha_i = f(x_i),  \quad \beta_i = h(x_i)
\end{equation}
where $f(x_i)$ and $h(x_i)$ represent functions of the data $x_i$ that may be implemented as neural networks; a method dubbed Dynamic Layer Normalisation when applied to an adaptive neuro acoustic model where the normalisation parameters were learned on the input sequence such that the model adapted to different input sequences \cite{Kim}.

\subsection{Feature-wise Linear Modulation}

An application of Dynamic Normalisation to visual reasoning \cite{Perez2018} demonstrated an RNN trained on input letter sequences that modified a CNN whose input was an image to answer queries such as "What number of cylinders are small purple things or yellow rubber things?". Dynamic normalisation was implemented by Feature-wise Linear Modulation (FiLM) layers conditioned on another input. The conditioning input was the output of an LSTM, adapting the CNN in response to textual user queries. FiLM layers act on intermediate features of the CNN.

With the FiLM function $F\in\mathbb{R}^d$, transforming the function $R\in\mathbb{R}^d$ with $\alpha\in\mathbb{R}^d$ and $\\\beta\in\mathbb{R}^d$ :

\begin{equation}
    F(R_{i,c}|\alpha_{i,c},\beta_{i,c})=\alpha_{i,c} \odot R_{i,c} + \beta_{i,c}
\end{equation}

\begin{equation}
    \alpha_{i,c} = f_c(x_i),  \quad \beta_{i,c} = h_c(x_i)
\end{equation}

where $R_{i,c}$ is the neural network activations for the i\textsuperscript{th} input and c\textsuperscript{th} feature map; $f_c(x_i)$ and $h_c(x_i)$ are the learned modulation factors that may be implemented as Neural Networks e.g. a Dense layer.

\subsection{Loss Conditional Training}\label{sec:Loss Conditional Training}
During optimisation of neural networks, loss functions often have to reflect multiple objectives, resulting in loss functions that are composed of a weighted sum of individual aspects of the loss. For example, a loss function for a Beta-Variational Autoencoder ($\beta$-VAE) \cite{Higgins2019}, balances reconstruction accuracy against the disentanglement of the latent representations. In artistic style transfer where an image is transformed to the style of another image, a weighted sum of losses from various layers are optimised, resulting in different image stylisation as the losses are weighted differently \cite{Babaeizadeh}.

Dynamic normalisation has been shown to be beneficial for adapting a neural network conditional on another network's output, but conditional normalisation is also applicable where the conditioning input may be a hyper-parameter of the network. Conditioning on the loss is possible where feature wise linear modulation \cite{Perez2018} was applied  to condition a neural network on the loss parameters such that a single trained model covers the whole weighting space of the losses \cite{Dosovitskiy}. 

Because the model is trained over the distribution of conditioned input values, a single model can operate during inference time at any operating point designated by the conditioning input without costly retraining. When conditioned on a loss hyper-parameter, a $\beta$-VAE trained for all $\beta$-values operated similarly to several models individually trained for each $\beta$-value \cite{Dosovitskiy}.

\section{Method}
\label{sec:method}
\subsection{Noise Conditional Training}\label{sec:Noise Conditional Training}
In this section conditional normalisation is applied to the problem of discriminative denoising, in which a fixed discriminative denoiser is generalised by conditioning on noise parameter inputs during training.

Consider a typical denoising problem where the task is to solve an inverse problem to reconstruct the ground truth images from the sensor images corrupted by noise. Given a training distribution of ground-truth, input image pairs $x,y\sim P_{x,y}$ with  $x \in X \subset \mathbb R^{d}$ and $y \in Y \subset \mathbb R^{d}$, forming a dataset $\mathcal {D}$ $$\mathcal{D}=\{(x_n,y_n)\}_{n=1}^N$$ such that $H:X \rightarrow Y$ transforms the ground truth image $x$ to the measured image $y$. We need to find a function $R$ to reconstruct the image $R:Y \rightarrow X$. Typically the dataset D comprises clean / noisy image pairs where the clean image may be corrupted to form the noisy image before being applied to the denoising network. Alternatively, in a scenario where ground truth data does not exist, a noise adding layer may corrupt an input image to obtain the noisy input.

We wish to train a Neural Network to find $R_{opt}$ :
\begin{equation}\label{equ:Ropt}
    R_{opt}=\arg\min_{R_\theta,\theta\in\Theta}\sum_{n=1}^{N}L(x_n,R_\theta(y_n))+g(\theta)
\end{equation}

where $R_\theta$ is the neural network function, $\theta$ represents the neural network weights, $\Theta$ represents the set of possible weights, $L:X \times X \to \mathbb{R}^+$ is the loss function and $g:\Theta \to \mathbb{R}^+$ is a regularisation function capturing prior knowledge of x. It may be noted that the structure of the neural network may serve as an implicit prior \cite{Lempitsky2018}.

The function R reconstructs a stochastic corruption of x:
\begin{equation}\label{equ:Corruption}
    y=N(x), \quad \text {where} \ N(x)=x+\epsilon
\end{equation}

Without limitation, we assume a Poisson-Gaussian noise model. Since Poisson noise is signal dependent, the Poisson-Gaussian noise model may be described by a generic signal dependent model:
\begin{equation}\label{equ:sig_noise}
    y=N(x)=x+\eta(x)\mathcal{N}(0,I)
\end{equation}
where y is the measured image, x is the ground-truth, noise free image, $\eta(x)$ is an image dependant standard deviation, that can be expressed as two terms, one dependant and one independent of the image:
\begin{equation}\label{equ:pg-std}
    \eta^2(x)=\alpha x + \sigma^2
\end{equation}
where $\alpha$ is the Poissonian noise parameter and $\sigma$ is the Gaussian standard deviation \cite{Lee2018,Foi2008}.

Typically, the loss is minimised for a specific value of $\sigma$ :
\begin{equation}
    \theta=\arg\min_\theta \mathbb{E}_x [ L(x,R(N(x,\sigma),\theta)]
\end{equation}

Instead of $\sigma$ being fixed, we propose to sample $\sigma$ from a distribution $P_\sigma$

Therefore the solution of the following optimisation problem is required:
\begin{equation}\label{equ:Loss_opt}
    \theta_{opt}=\arg\min_{\theta} \mathbb{E}_\sigma [\mathbb{E}_x[ L(x,R(N(x,\sigma),\theta,\sigma))]]
\end{equation}

Employing stochastic gradient descent to optimise (\ref{equ:Loss_opt}) with  Monte Carlo estimates of the expectations over the training set D and $\sigma_i \sim P_{\sigma}$ :
\begin{equation}
    \theta=\arg\min_{\theta} \sum_{i=1}^n L(x_i,R(N(x_i,\sigma_i),\theta,\sigma_i))
\end{equation}

Conditioning the denoising neural network R on the distribution of noise corruption $\sigma_i \sim P_{\sigma}$  is achieved by adding FiLM layers \cite{Perez2018} to the denoising neural network.

With the FiLM function $F\in\mathbb{R}^d$, transforming the function $R\in\mathbb{R}^d$ with $\alpha\in\mathbb{R}^d$ and $\\\beta\in\mathbb{R}^d$ :

\begin{equation}\label{equ:film_1}
    F(R_{i,c}|\alpha_{i,c},\beta_{i,c})=\alpha_{i,c} \odot R_{i,c} + \beta_{i,c}
\end{equation}

\begin{equation}\label{equ:film_2}
    \alpha_{i,c} = f_c(x_i),  \quad \beta_{i,c} = h_c(x_i)
\end{equation}

where $R_{i,c}$ is the neural network activations for the i\textsuperscript{th} input and c\textsuperscript{th} feature map; $f_c(x_i)$, $h_c(x_i)$ are the learned modulation factors and $\odot$ signifies the Hadamard product.

\subsection{Implementation}\label{sec:NoiseCondImplementation}

We have chosen U-Net as the neural network model for implementation of Noise Conditional training, since U-Net \cite{Ronneberger2015} has been shown to act an implicit prior on the structure of natural images \cite{Lempitsky2018}, aiding in the regularisation of equation (\ref{equ:Ropt}), and has been demonstrated to perform well as a denoiser \cite{Chen2018,Jin2017a}.

\begin{figure}
    \centering
    \includegraphics[width=8cm,keepaspectratio]{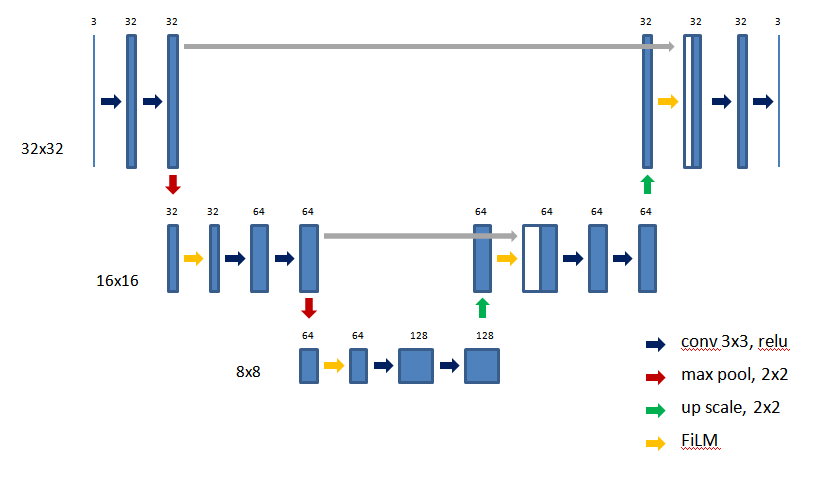}
    \caption{U-Net with FiLM Layers (32,32,3)}
    \label{fig:UNet_Film_32323}
\end{figure}

The model architecture is illustrated in Figure \ref{fig:UNet_Film_32323}, configured with a (32,32,3) dimensional input layer to process full size RGB images from the CIFAR-10 dataset \cite{Krizhevsky2009} and image patches from larger images. A noise layer is added prior to the input, implementing the corruption process of equations (\ref{equ:sig_noise},\ref{equ:pg-std}) so the input image  serves also as the ground truth in the same manner as a denoising autoencoder. The Film layers implement equations (\ref{equ:film_1},\ref{equ:film_2}), with the Poisson-Gaussian noise layer parameters ($\alpha$, $\sigma$) as the conditioning inputs.

\section{Experiments}
\label{sec:experiments}

\subsection{Comparison of U-Net and Noise Conditional U-Net}
The Film-UNet model was trained on 45000 examples of the CIFAR-10 dataset, with a corresponding number of samples from a uniform distribution $P_\sigma \sim \sigma \in [0,1]$, $\alpha=0$. The CIFAR-10 images were applied to U-Net, with $P_{\sigma}$ applied to the conditioning input. The network was trained with an MSE loss function:
\begin{equation}
    L=\frac{1}{N}\sum_{i=1}^{N}(y_i - x_i)^2
\end{equation}
The optimiser was ADAM \cite{Dietterich2017} (learning rate=0.001, epsilon=1E-7; the default parameters of Tensorflow 2). The batch size was 64.

Once trained, the model was validated on 5000 CIFAR-10 example images with various values of $\sigma$ (corresponding to sampling from $P_\sigma$), being presented to the conditioning input. The noise adding layer is active only during training. A comparison was performed between U-Net trained several times on various noise corruption values, and Film-Unet where the noise level was presented to the conditioning input at inference time.

\begin{figure}
    \centering
    \includegraphics[width=8cm,keepaspectratio]{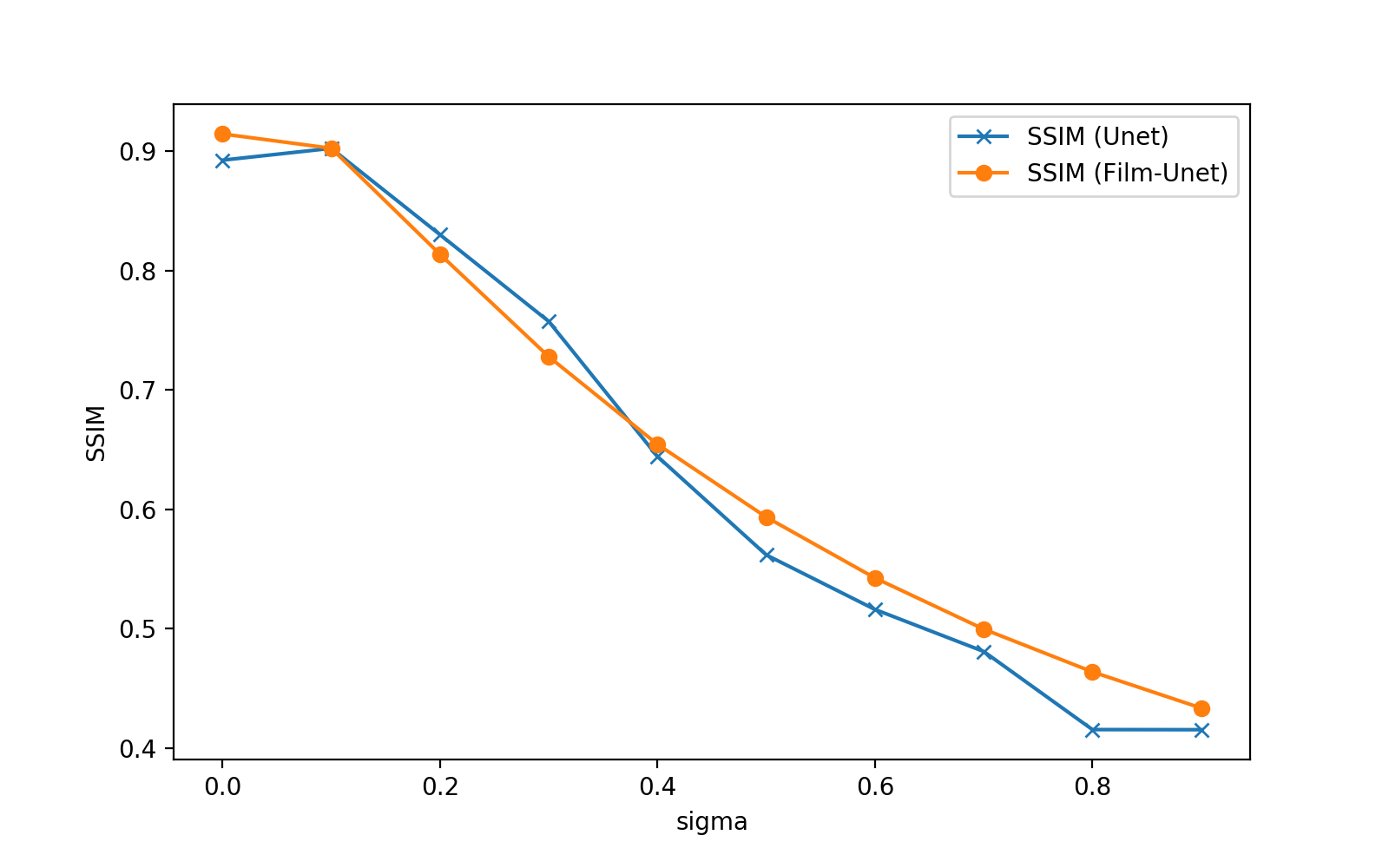}
    \caption{SSIM versus $\sigma_{tr}$ ($\sigma_{val}=0.05$)}
    \label{fig:SSIM_sig_UNet_Film_32323}
\end{figure}

 Mean SSIM comparisons are shown in Figure \ref{fig:SSIM_sig_UNet_Film_32323}. For validation the images are corrupted with Gaussian noise $\epsilon = \mathcal{N}(0,\sigma_{val}^2) = \mathcal{N}(0, 0.05^2)$. The results show that the U-Net model trained conditionally on Gaussian noise over a noise level distribution $\sigma \sim P_{\sigma_{tr} \in [0,1]}$,  performs similarly to several U-Net models trained independently for various levels of corruption noise during training.
 The training time for the noise conditional model was 8 minutes for all noise levels, versus 7 minutes for each noise level on the standard model (RTX2070S GPU). For example taking 10 noise levels, training time is reduced from 70 minutes to 8 minutes; a significant improvement.
 
\subsection{A Noise Conditional Model on Poisson-Gaussian Noise}
A  Film-Unet model was trained over distribution ($\alpha,\sigma)\in([0,1],[0,1]$). The models were then validated on both Gaussian and Poisson noise to understand the performance impact that the model training noise has on the image noise distribution during validation.  The experiments were run using the CIFAR-10 dataset using 45000 training images and 5000 validation images. 

\begin{figure}[t]
    \centering
    \includegraphics[width=8cm,keepaspectratio]{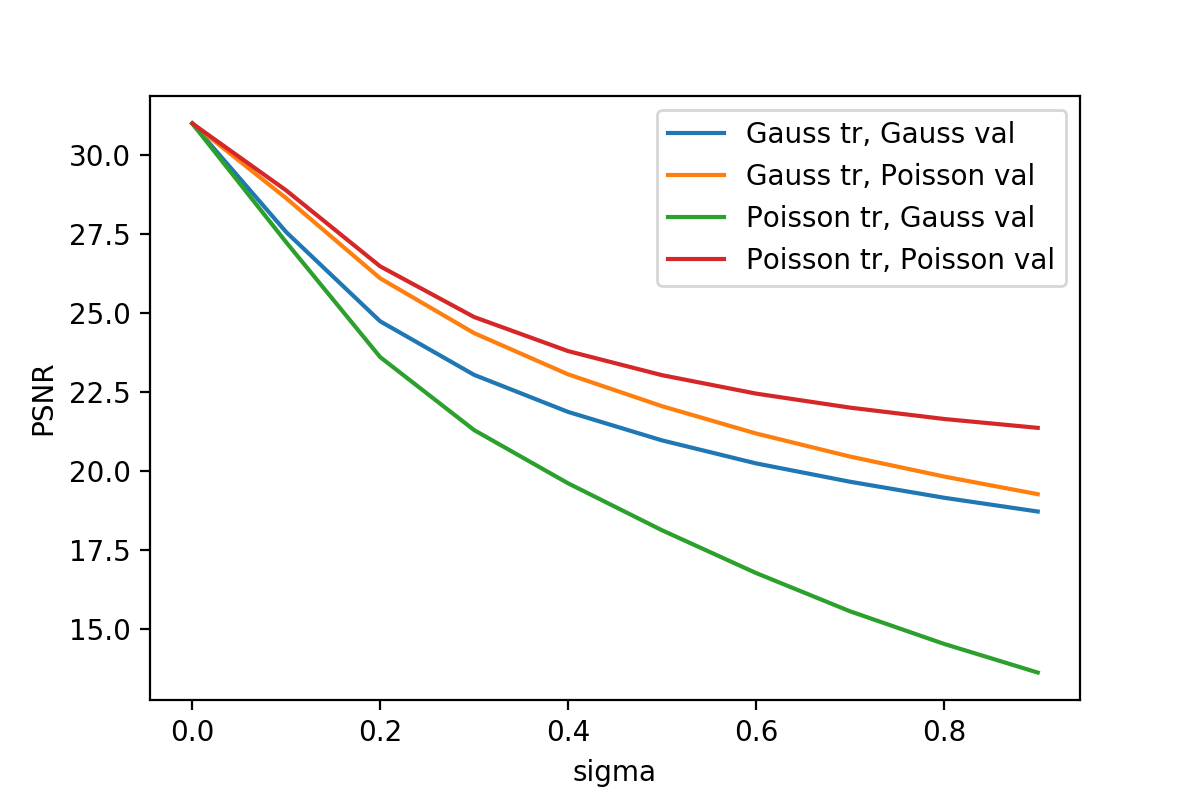}
    \caption{PSNR curves versus $\sigma_{val}$ }
    \label{fig:PSNR_CIFAR10_PG_FILMUnet}
\end{figure}

A model trained on a distribution of Gaussian noise $\sigma_{tr}\in[0,1]$ was validated with $\sigma_{val}\in[0,1]$ at points $\sigma\in\sigma_{tr}=\sigma_{val}$, where $\sigma$ corresponds to the Gaussian noise standard deviation, $\sigma$ of equation (\ref{equ:pg-std}). The PSNR curves are shown in Figure \ref{fig:PSNR_CIFAR10_PG_FILMUnet}. The yellow and blue curves show that training on Gaussian noise produces similar performance for both Gaussian and Poisson validation noise. However, training on Poisson noise considerably improves the performance of Poisson input noise during validation (red curve), whilst Gaussian noise performance (green curve) is reduced. This indicates that training with Poisson noise corruption improves over Gaussian training on Poisson noise corrupted images.

\subsection{Two Phase Training}
In this section we examine whether a pretext learning task with fixed noise-model parameters may be generalised in a 2nd phase of training, and whether this confers any advantages.

The model U-Net layers may be pretrained on fixed noise parameters in a first phase and subsequently the Film layers trained conditionally over a distribution of noise parameters. 

The role of the first phase is to train U-Net to denoise images with a fixed noise level applied. The training noise ($\sigma_{tr}$=1.0) is chosen to maximise the denoising in the first training phase.  The Film layers are present in the network but are not trained, have random initialisations, with the conditioning input at zero. This phase of training involves 764,323 total parameters, of which 518,563 are trained and 245,760 are not trained.

The Film layers are trained in the second phase, conditioned on $\sigma\in([0,1])$, with the U-Net layers being fixed, retaining the weights learned from the first phase. This phase of training involves 764,323 total parameters, of which 245,760 are trained and 518,563 not trained. 

\begin{figure}
    \centering
    \includegraphics[width=8cm,keepaspectratio]{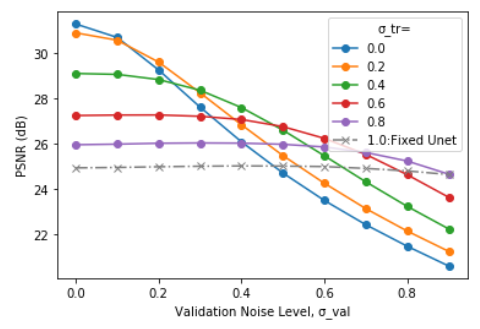}
    \caption{PSNR curves versus $\sigma_{val}$ (pretext $\sigma_{tr}=1.0$)}
    \label{fig:Pretext_2ph_PSNR_CIFAR_Film_32323}
\end{figure}

The performance following the 2nd phase of training is illustrated by the PSNRcurves shown in Figure \ref{fig:Pretext_2ph_PSNR_CIFAR_Film_32323}. The curves show results for the Film-Unet over a set of conditional input values $\sigma_{tr}\in{0,0,2,0,4,0.6,0.8}$. The performance of Unet trained at $\sigma_{tr}=1.0$ in the pretext phase is shown as a grey dotted curve.

It is clear from these results that a fixed Unet densoiser, utilising only the Unet layers of Film-Unet may be generalised to a range of noise levels by training the Film layers in a second training phase.

\subsection{Validation on Large Images}
Having developed and validated a noise conditionally trained Film-Unet on small images from the CIFAR10 dataset, this section seeks to validate the model on full size images. The training set comprises 70 medium exposure images and 20 validation images from the S7 dataset with 12M pixel resolution (3024,4032,3) \cite{Schwartz2018}. The CBSD68 dataset was also used for validation \cite{Martin}.

The Film-Unet model consists of 16,004,131 trainable parameters, with an input patch size of (128,128,3).

A single Film-Unet model was trained with Poisson-Gaussian noise (equation \ref{equ:pg-std}) over distribution ($\alpha,\sigma)\in([0,0.3],[0,0.3]$), similar to other studies; e.g.  \cite{Plotz2017} states $\sigma=0.01 \in [0,1]$ or $\sigma=2.7 \in [0,255]$ as realistic for cameras, whilst many other studies examine up to $\sigma=0.2 \in [0,1]$ or $\sigma=50 \in [0,255]$ \cite{Zhang2018}.

The Film-Unet was trained for 300 epochs with an MSE loss function and ADAM optimiser (Huber loss was also tried since it is more robust to high levels of noise but results were similar and are excluded for brevity). Training time was approximately 12 hours on an Nvidia RTX2070S GPU.

\subsection{Sensitivity Curves: CBSD68 Dataset}
For validation of the sensitivity curves, the PSNR, SSIM and output noise were measured on the validation set with various levels of Gaussian noise ($\sigma_{val}$) added. The training set consisted of 110 images from the S7 dataset, trained on (128,128,3) patches. Validation was on the CBSD68 dataset. PSNR, SSIM and output noise estimates were based on 640 (128,128,3) image patches.

As a whole, these curves show that denoising is effective up to the validation noise level corresponding to the training noise level, as would be expected.

\begin{figure}
    \centering
    \includegraphics[width=8cm,keepaspectratio]{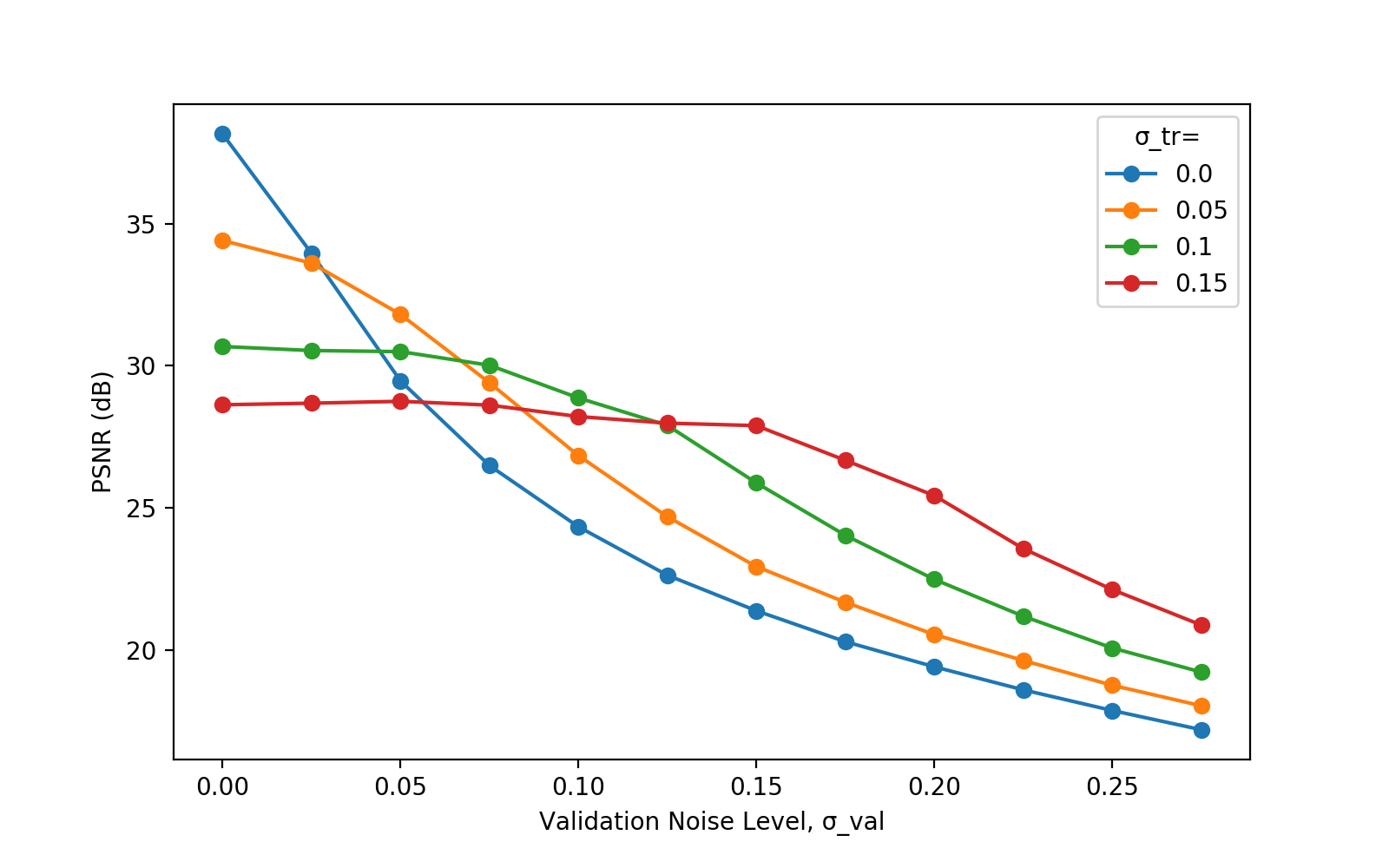}
    \caption{PSNR of Film-Unet with Gaussian Validation Noise}
    \label{fig:PSNR_BSD68_Film}
\end{figure}

\begin{figure}[t]
    \centering
    \includegraphics[width=8cm,keepaspectratio]{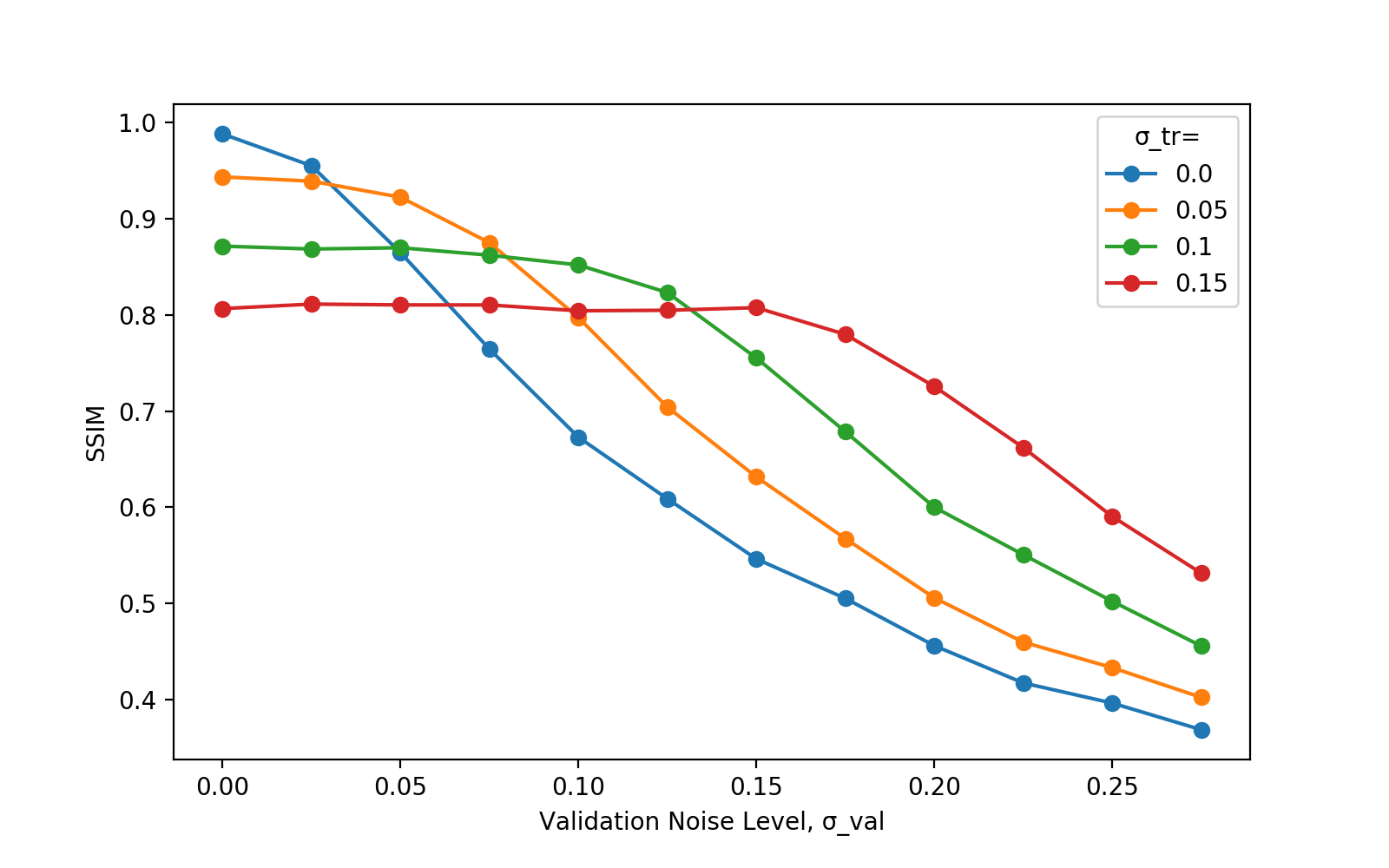}
    \caption{SSIM of Film-Unet with Gaussian Validation Noise }
    \label{fig:SSIM_BSD68_Film}
\end{figure}

The PSNR curves (Figure \ref{fig:PSNR_BSD68_Film}), are approximately flat until the validation noise level ($\sigma_{val}$), corresponding to the conditioning  noise level ($\sigma_{tr}$) for the curve is reached, after which the PSNR drops off. Over the family of curves, the maximum PSNR value for each $\sigma_{val}$ corresponds to the curve conditioned on the same value; i.e. ($\sigma_{tr}$,$\sigma_{val}$)=(0.2,0.2) etc. The SSIM curves (Figure \ref{fig:SSIM_BSD68_Film}) show a similar pattern.

\subsection{Comparison with other Methods}

BM3D (or CMB3D) has been shown to out perform many other methods on real images \cite{Plotz2017} and we considered it to be  the main algorithm of comparison in this paper for realistic images. FFDNet \cite{Zhang2018} is considered to be a leading example of CNN based image denoising \cite{Fan2019} and is also compared against on certain tests. A CBM3D Python Tensorflow implementation was used for these experiments \cite{Makinen2020}, and a Python implementation of FFDNet \cite{Zhang2020}.

\begin{table}[t]
\centering
\caption{PSNR Comparison CBSD68 and S7 datasets}
\label{tab:PSNR Comparison}
\begin{tabular}{|l|l|l|l|l|l|l|}
\hline
Method & $\sigma=0.05$ & $\sigma=0.10$ & $\sigma=0.20$ & $\sigma=0.30$ \\ \hline
Gaussian & CBSD68 &&& \\ \hline
CBM3D     & 31.2        & 27.8      & 26.3      & 23.5        \\
FFDNET    & 32.7        & 28.6      & 26.4      & 21.9       \\
Film-Unet & 31.6        & 29.0      & 27.7      & 25.2        \\ \hline
Poisson & CBSD68 &&& \\ \hline
CBM3D     & 31.4        & 27.9      & 25.3        & 24.0        \\
FFDNET    & 32.8        & 28.6      & 24.9        & 21.9        \\
Film-Unet & 34.0        & 31.3      & 28.5        & 27.0        \\ \hline 
Gaussian & S7 &&& \\ \hline
CBM3D     & 32.3        & 30.0      & 28.9      & 27.2        \\
Film-Unet & 33.6        & 31.4      & 30.4      & 29.0        \\ \hline
Poisson & S7 &&& \\ \hline
CBM3D     & 32.2        & 30.0      & 28.5        & 27.8        \\
Film-Unet & 35.6        & 33.1      & 31.1        & 30.2        \\ \hline
\end{tabular}
\end{table}

\begin{table}[]
\centering
\caption{SSIM Comparison CBSD68 and S7 datasets}
\label{tab:SSIM Comparison}
\begin{tabular}{|l|l|l|l|l|l|}
\hline
Dataset & $\sigma=0.05$ & $\sigma=0.10$  & $\sigma=0.20$ & $\sigma=0.30$ \\ \hline
Gaussian & CBSD68 &&& \\ \hline
CBM3D     & 0.88        & 0.78      & 0.72     & 0.60        \\
FFDNet    & 0.92        & 0.82      & 0.75     & 0.63        \\
Film-Unet & 0.92        & 0.86      & 0.82     & 0.71        \\ \hline
Poisson & CBSD68 &&& \\ \hline
CBM3D     & 0.87        & 0.77      & 0.66        & 0.61        \\
FFDNet    & 0.92        & 0.82      & 0.72        & 0.63        \\
Film-Unet & 0.96        & 0.91      & 0.84        & 0.79        \\ \hline
Gaussian & S7 &&& \\ \hline
CBM3D     & 0.83        & 0.75      & 0.71     & 0.67        \\
Film-Unet & 0.89        & 0.83      & 0.78     & 0.73        \\ \hline
Poisson & S7 &&& \\ \hline
CBM3D     & 0.81        & 0.73      & 0.69        & 0.67       \\
Film-Unet & 0.93        & 0.88      & 0.81        & 0.78        \\ \hline
\end{tabular}
\end{table}

Film-Unet is compared with other methods under Gaussian and Poisson noise in Tables \ref{tab:PSNR Comparison},\ref{tab:SSIM Comparison}, with both CBSD68 and S7 (uncompressed) used for validation. Each PSNR and SSIM metric evaluation was based on 40 full images from the dataset. For CMB3D the whole image was applied to the algorithm, for Film-Unet the image was divided into 128x128 patches, applied to the denoiser and re-assembled from denoised patches. The evaluation images had the required noise level and type pre-applied, so that identical images were applied to each method for comparison. The comparison with CBM3D in Tables \ref{tab:PSNR Comparison} and \ref{tab:SSIM Comparison}, shows that Film-Unet performs better at almost all validation noise levels and on both PSNR and SSIM metrics for both validation sets and noise types. For FFDNet on CBSD68 dataset with Gaussian noise, PSNR at $\sigma=0.05$ is better for FFDNet, whilst SSIM  at $\sigma=0.05$ is the same. Under Poisson Noise the performance of CBM3D and FFDNet does not deteriorate significantly compared to Gaussian noise, however, Film-Unet performance improves under Poisson noise since it has the benefit of having been trained on a PG noise model.

\begin{figure*}[t]
\caption{Detail of DND image 37. Poisson Noise. $\sigma_{val}=0.05$ (left),  $\sigma_{val}=0.10$ (middle), $\sigma_{val}=0.20$ (right). Input (bottom row), BM3D (2nd row), FFDNet (3rd row), Film-Unet (top row).}
\label{fig:DND 37_P}
\begin{tikzpicture}
\node at (0,0) {\includegraphics[width=6cm]{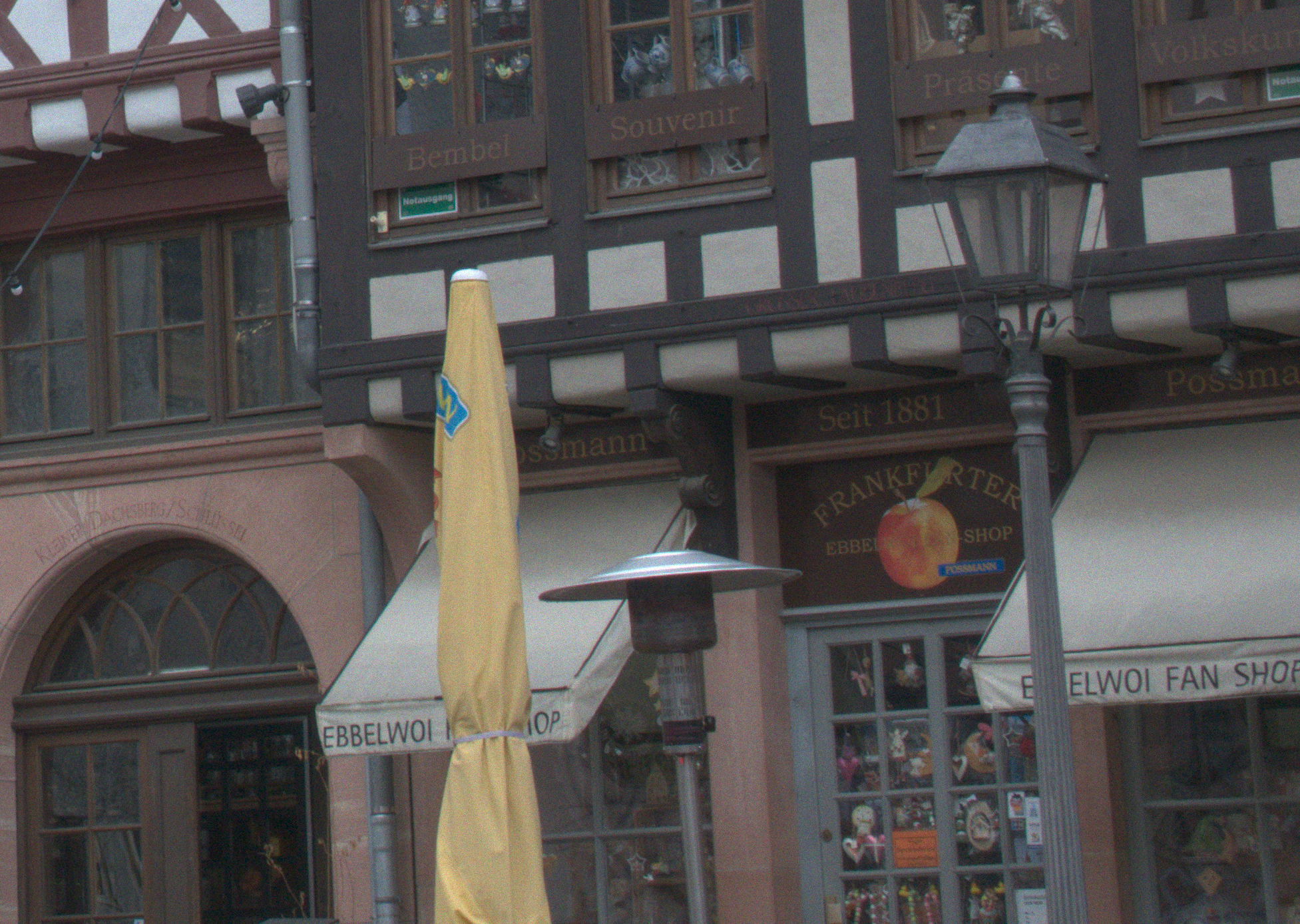}};
\node at (6,0) {\includegraphics[width=6cm]{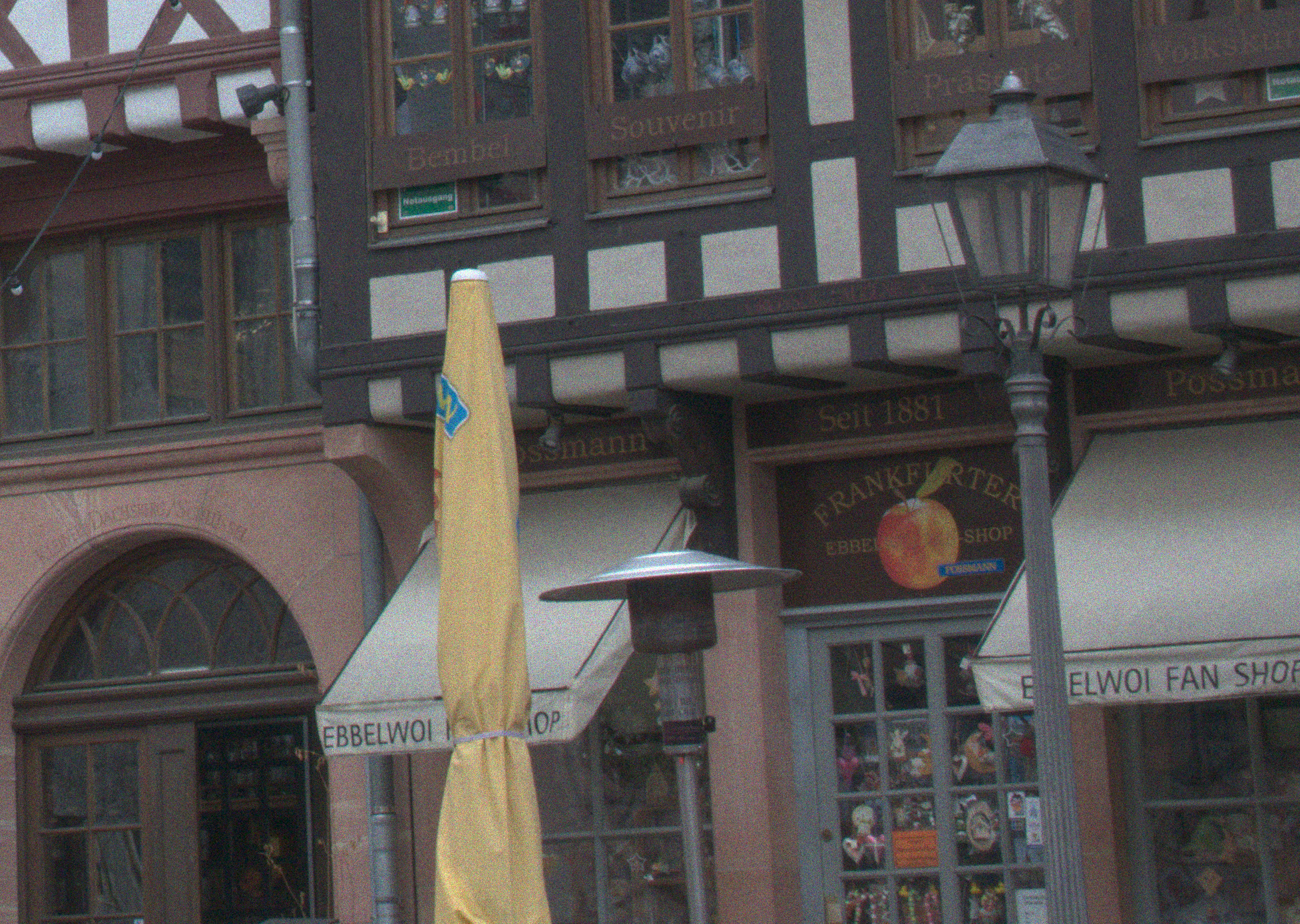}};
\node at (12,0) {\includegraphics[width=6cm]{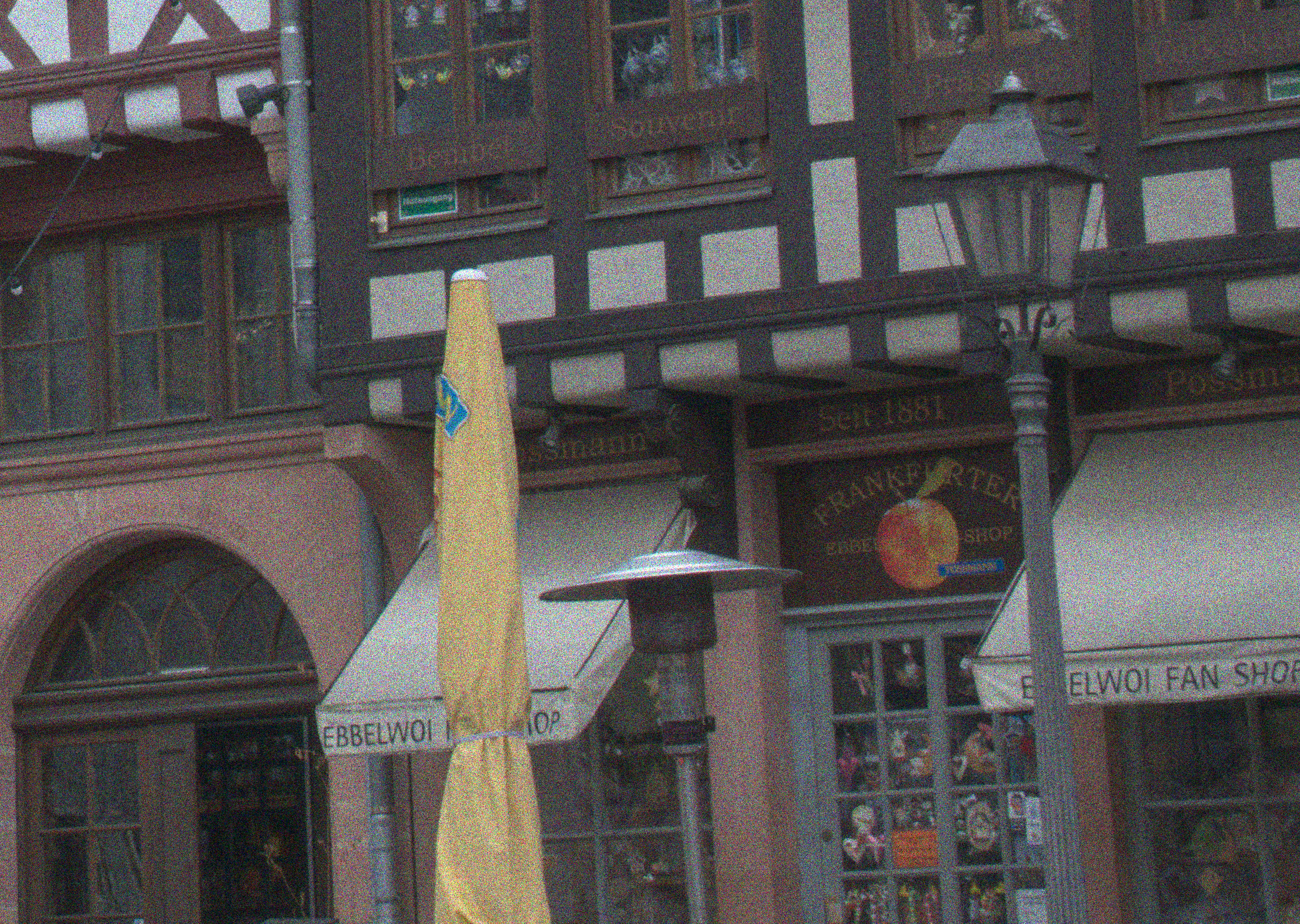}};
\node at (0,3.1) {\adjustbox{trim=\dimexpr\Width-6cm 1cm 0cm \dimexpr\Height-4cm \relax{} 0pt,clip}{\includegraphics[width=20cm]{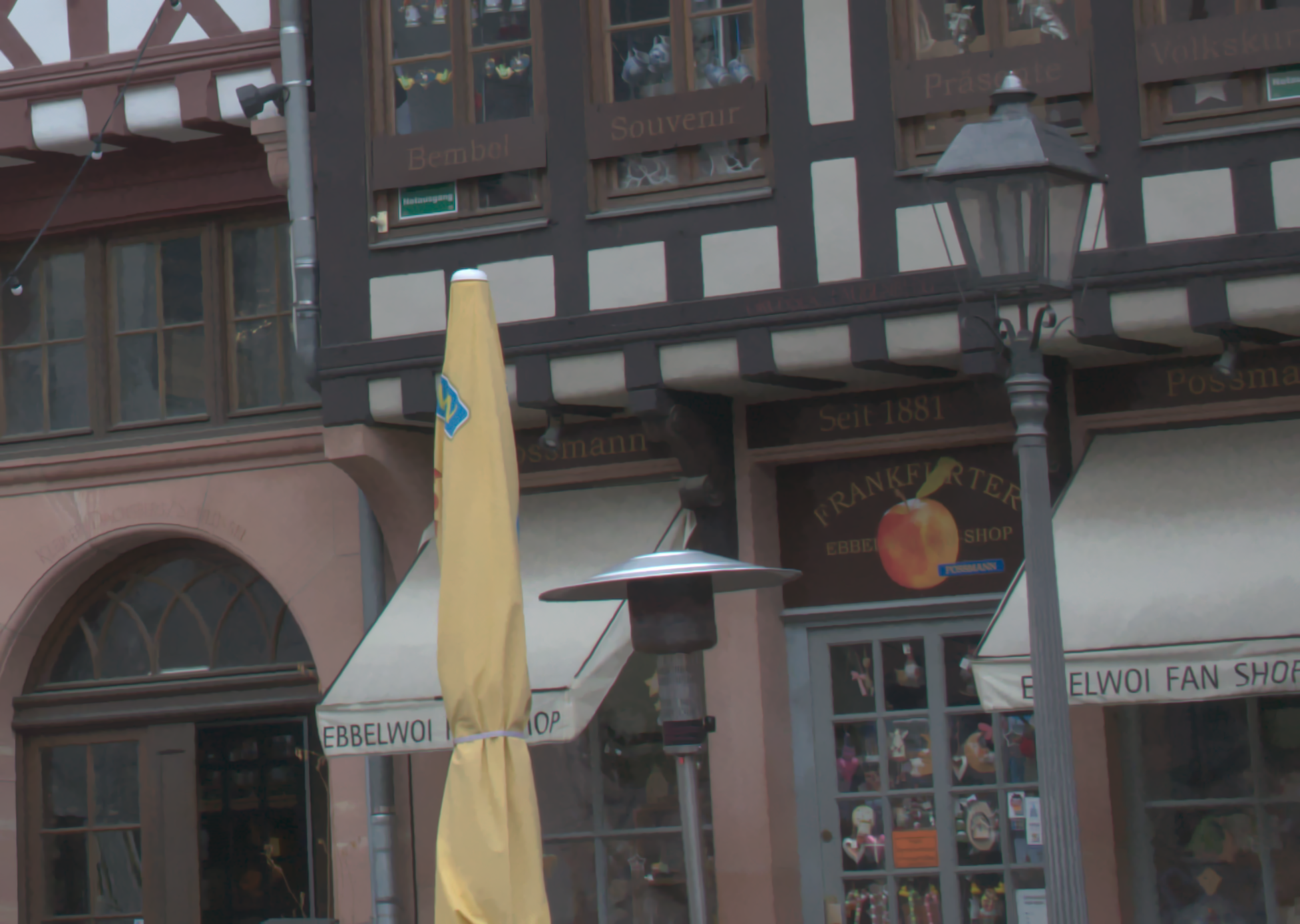}}};
\node at (6,3.1) {\adjustbox{trim=\dimexpr\Width-6cm 1cm 0cm \dimexpr\Height-4cm \relax{} 0pt,clip}{\includegraphics[width=20cm]{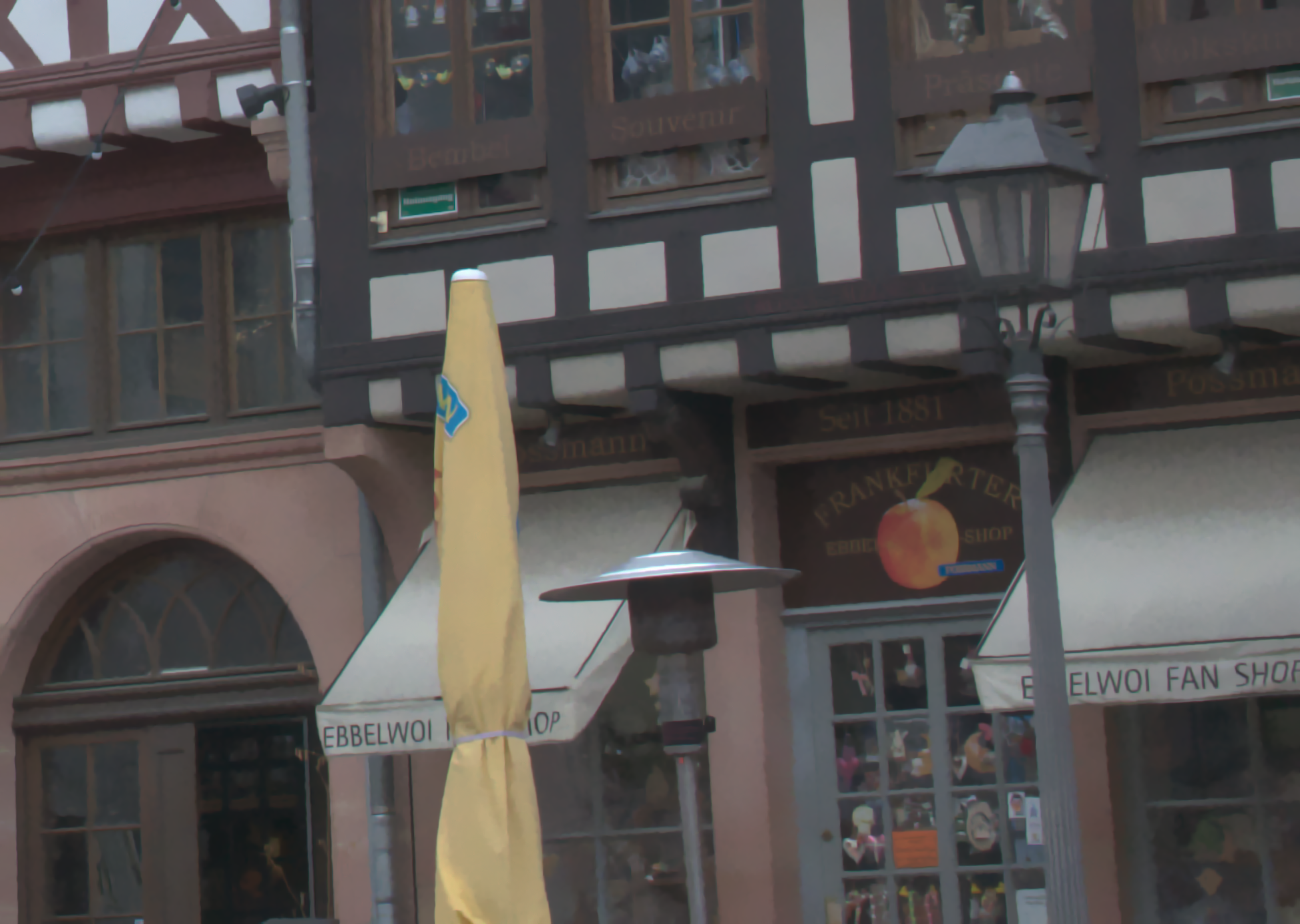}}};
\node at (12,3.1) {\adjustbox{trim=\dimexpr\Width-6cm 1cm 0cm \dimexpr\Height-4cm \relax{} 0pt,clip}{\includegraphics[width=20cm]{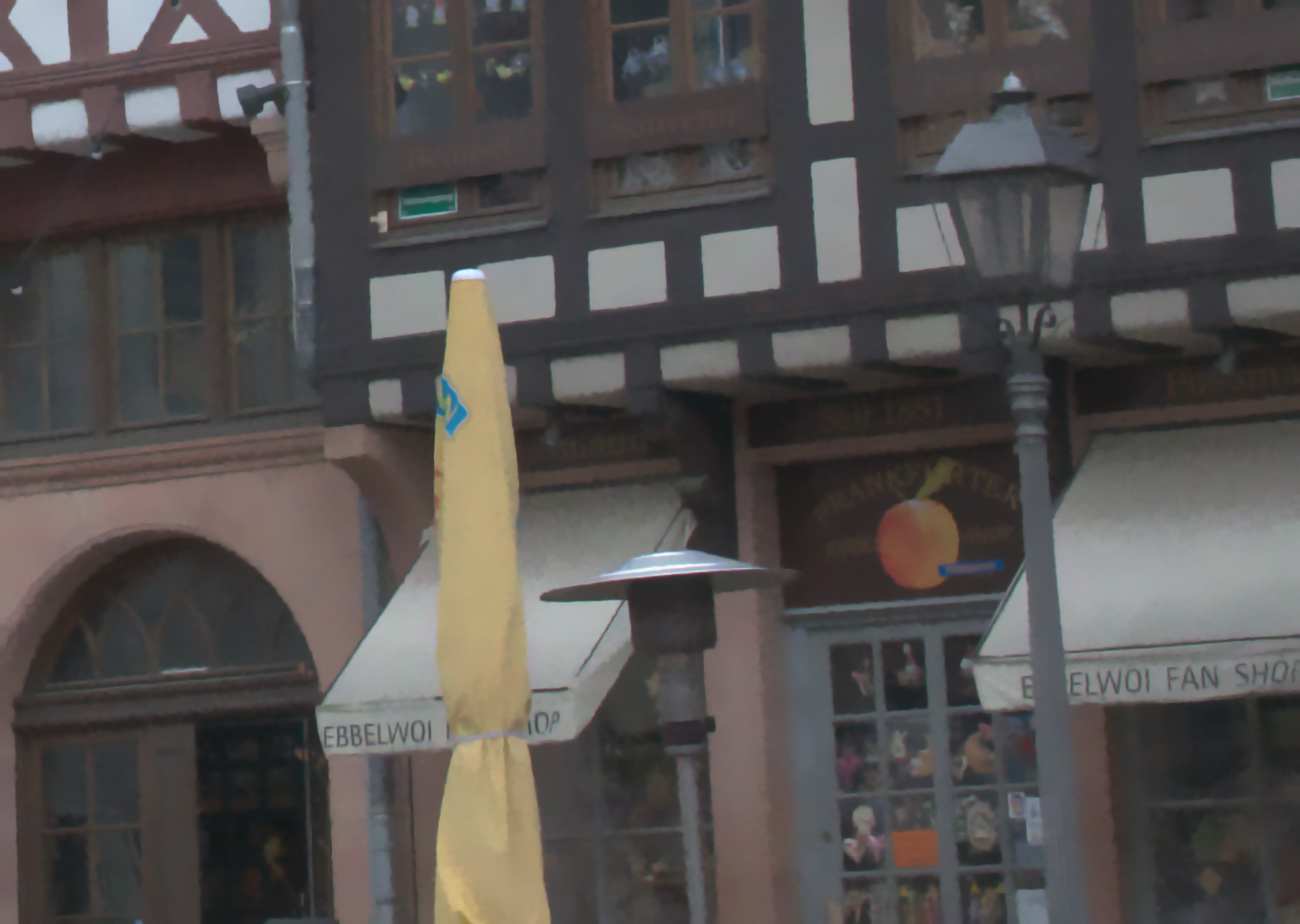}}};

\node at (0,6.1) {\adjustbox{trim=\dimexpr\Width-6cm 1cm 0cm \dimexpr\Height-4cm \relax{} 0pt,clip}{\includegraphics[width=20cm]{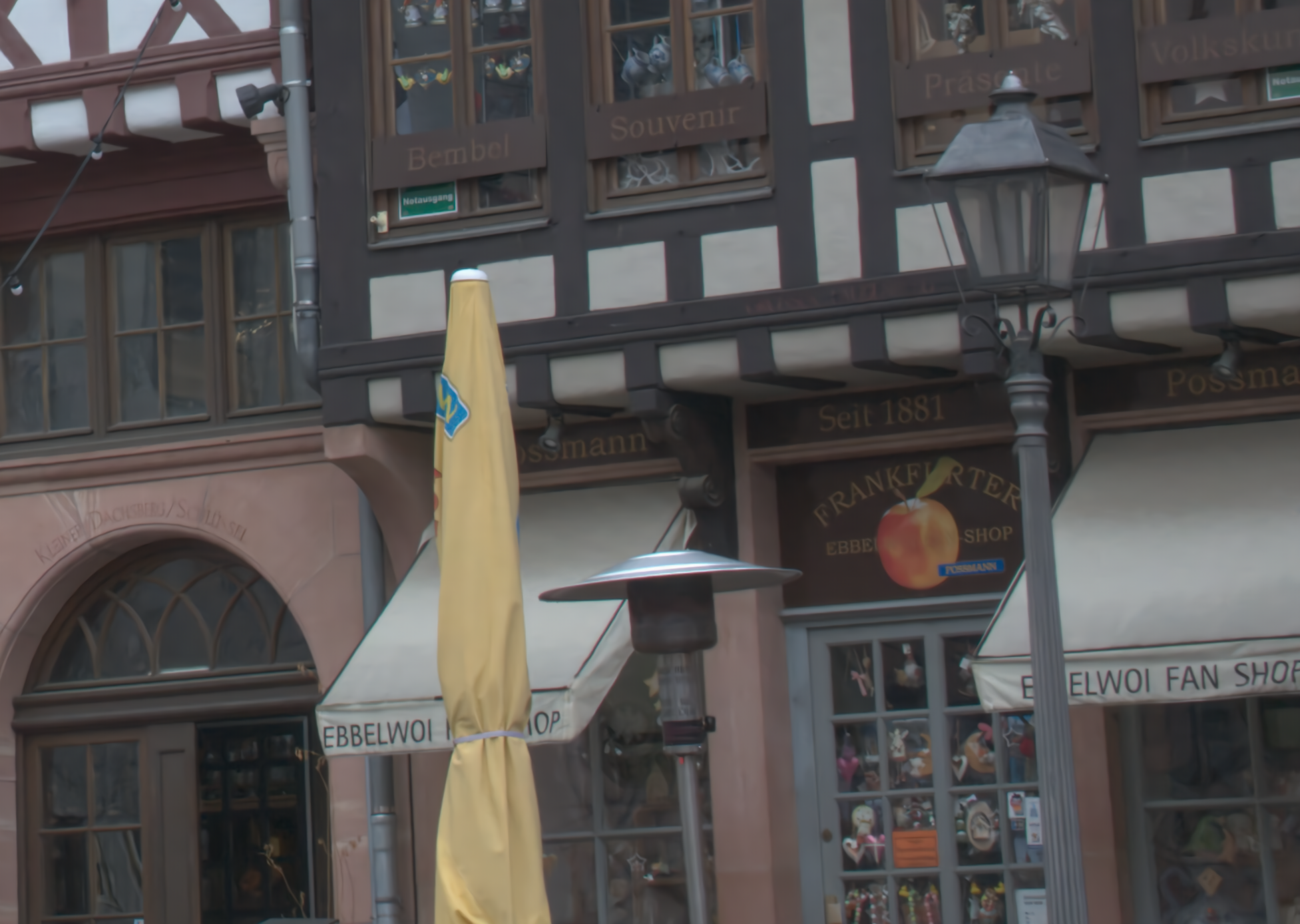}}};
\node at (6,6.1) {\adjustbox{trim=\dimexpr\Width-6cm 1cm 0cm \dimexpr\Height-4cm \relax{} 0pt,clip}{\includegraphics[width=20cm]{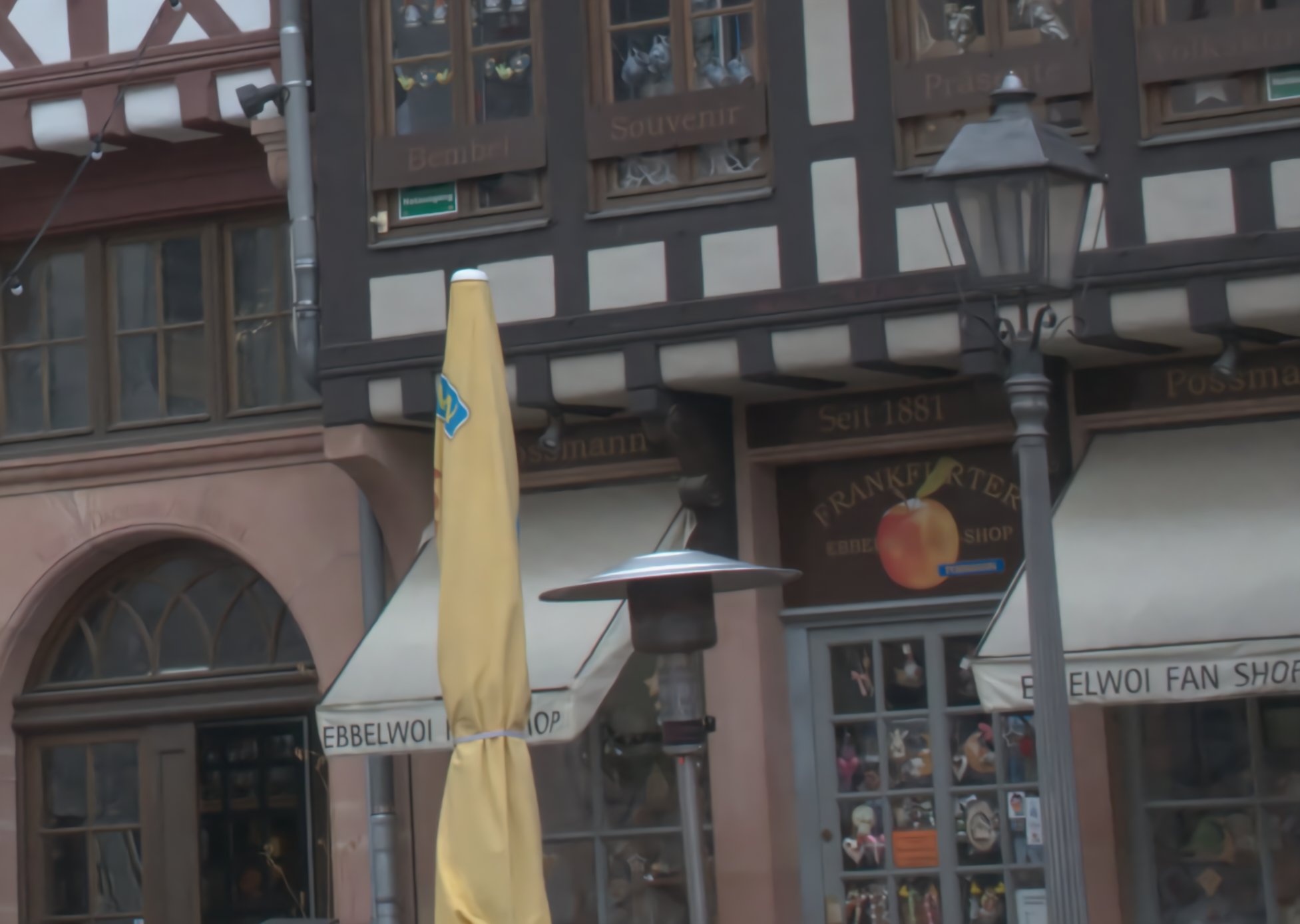}}};
\node at (12,6.1) {\adjustbox{trim=\dimexpr\Width-6cm 1cm 0cm \dimexpr\Height-4cm \relax{} 0pt,clip}{\includegraphics[width=20cm]{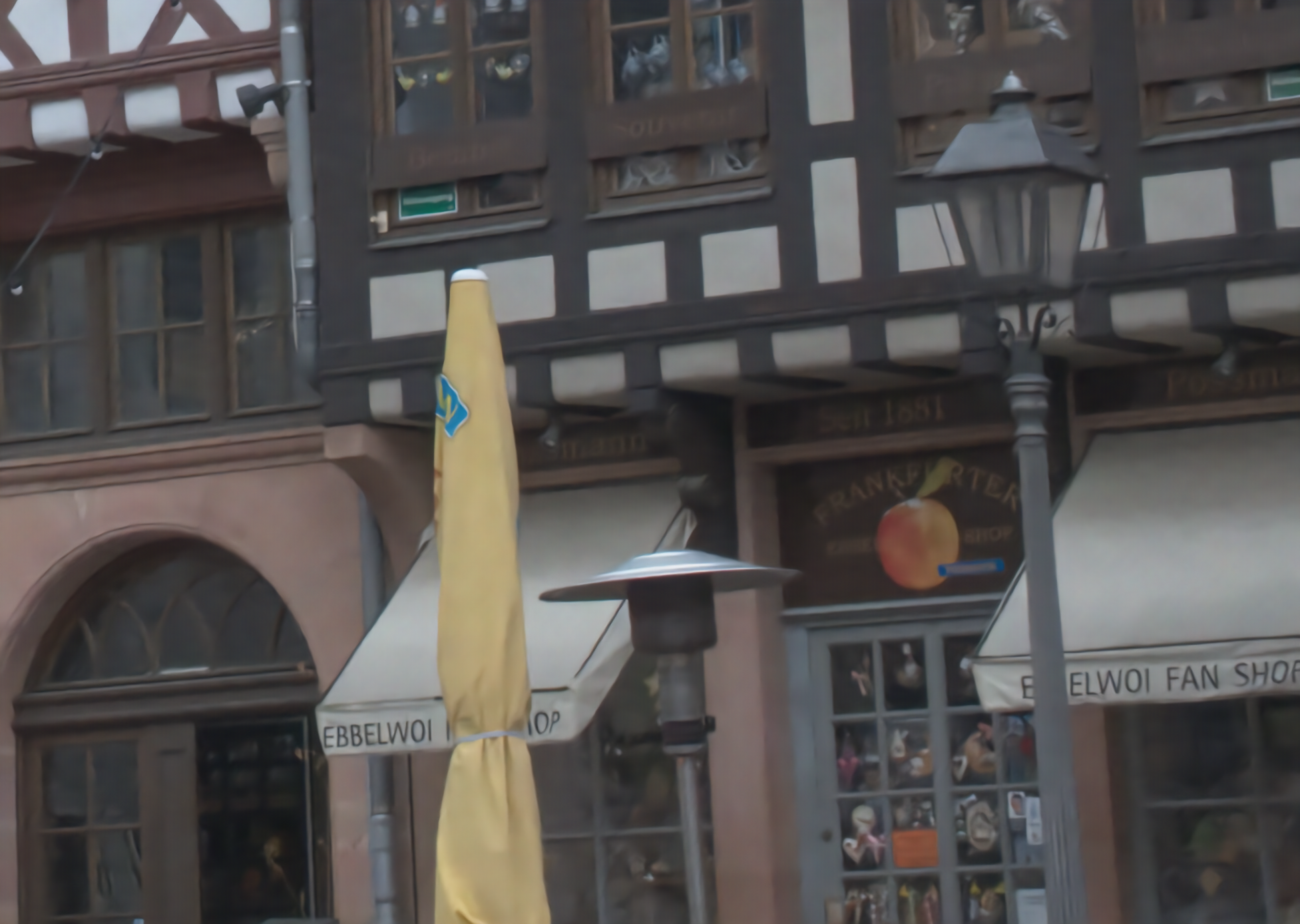}}};

\node at (0,9.1) {\adjustbox{trim=\dimexpr\Width-6cm 1cm 0cm \dimexpr\Height-4cm \relax{} 0pt,clip}{\includegraphics[width=20cm]{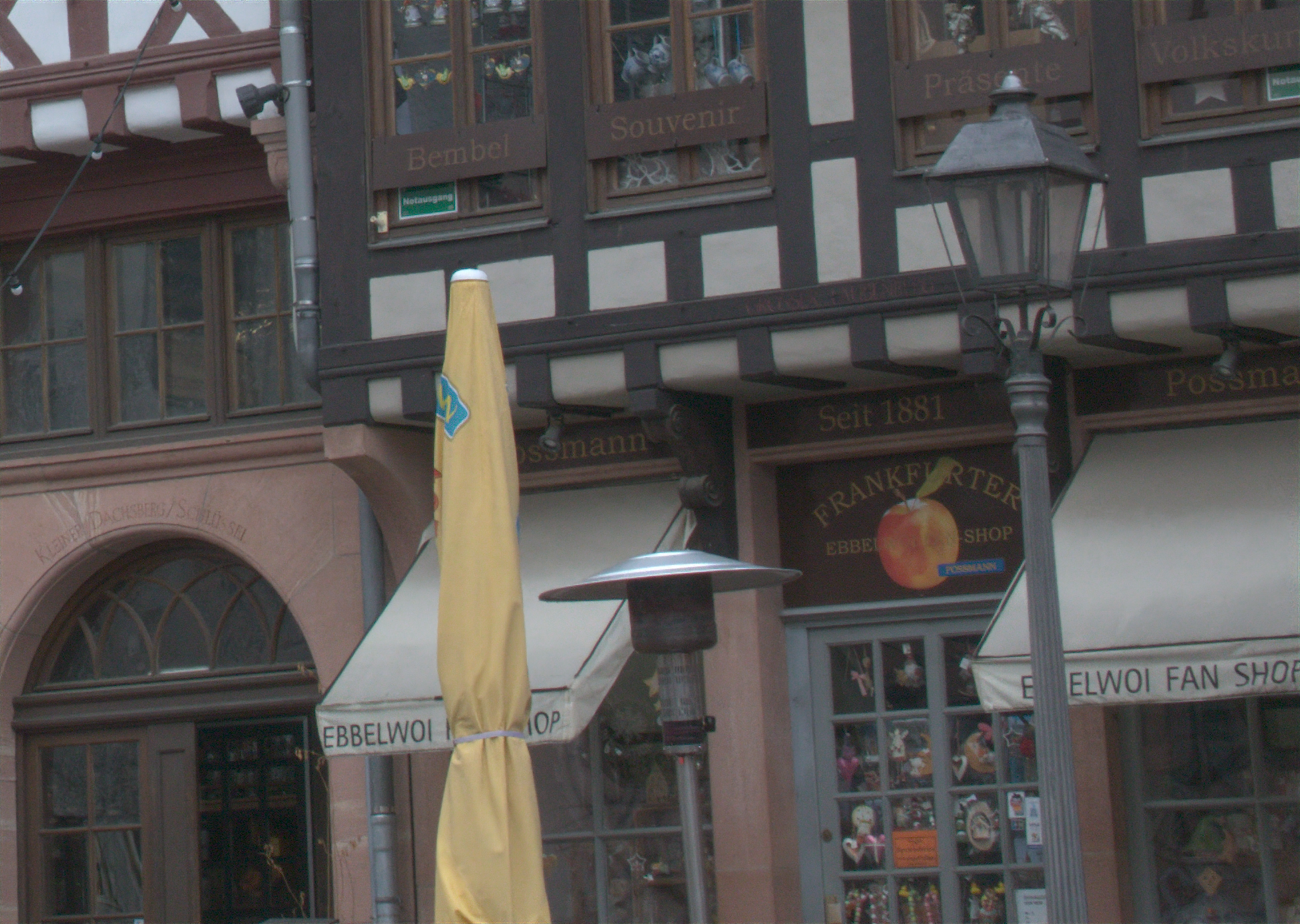}}};
\node at (6,9.1) {\adjustbox{trim=\dimexpr\Width-6cm 1cm 0cm \dimexpr\Height-4cm \relax{} 0pt,clip}{\includegraphics[width=20cm]{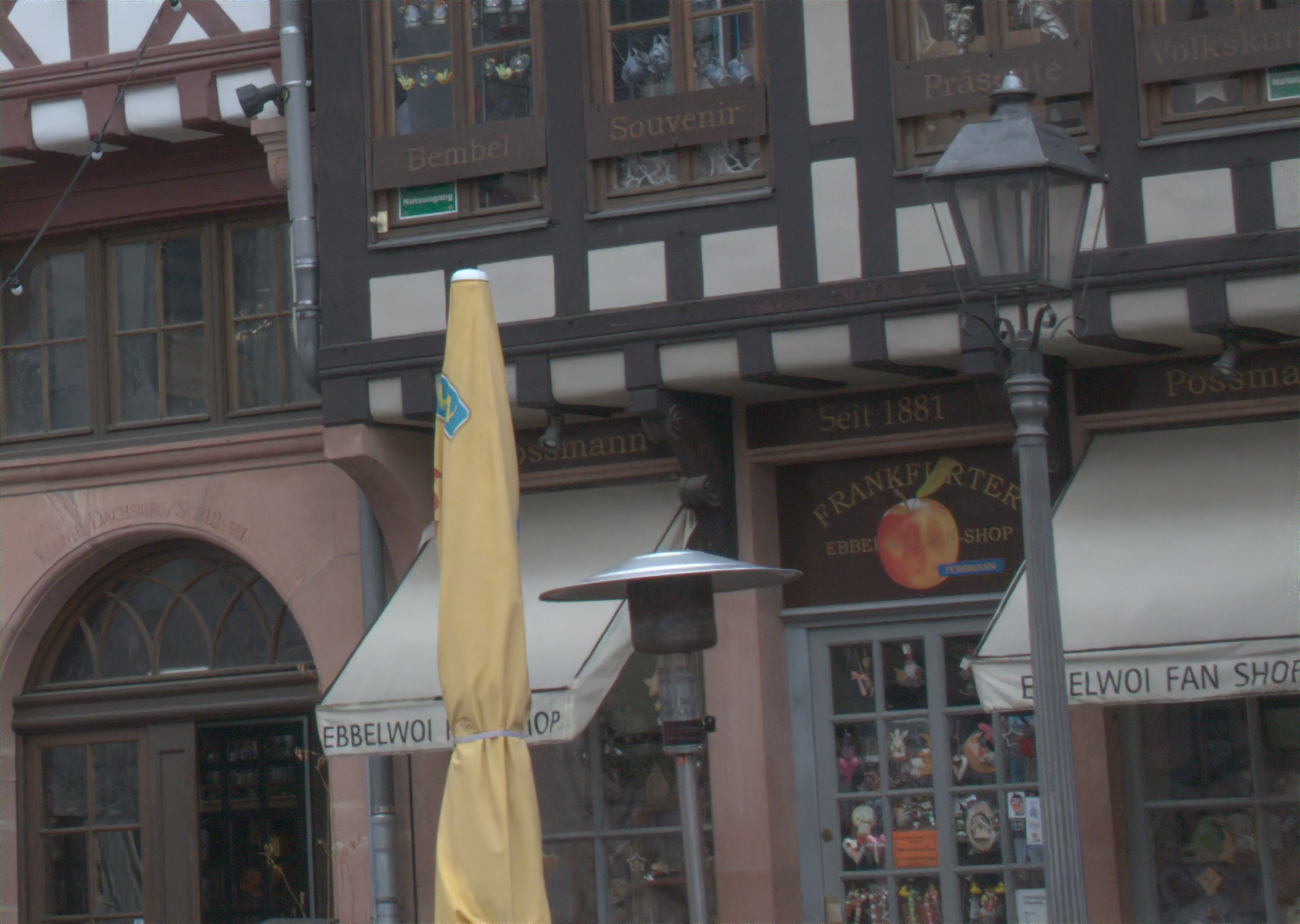}}};
\node at (12,9.1){\adjustbox{trim=\dimexpr\Width-6cm 1cm 0cm \dimexpr\Height-4cm \relax{} 0pt,clip}{\includegraphics[width=20cm]{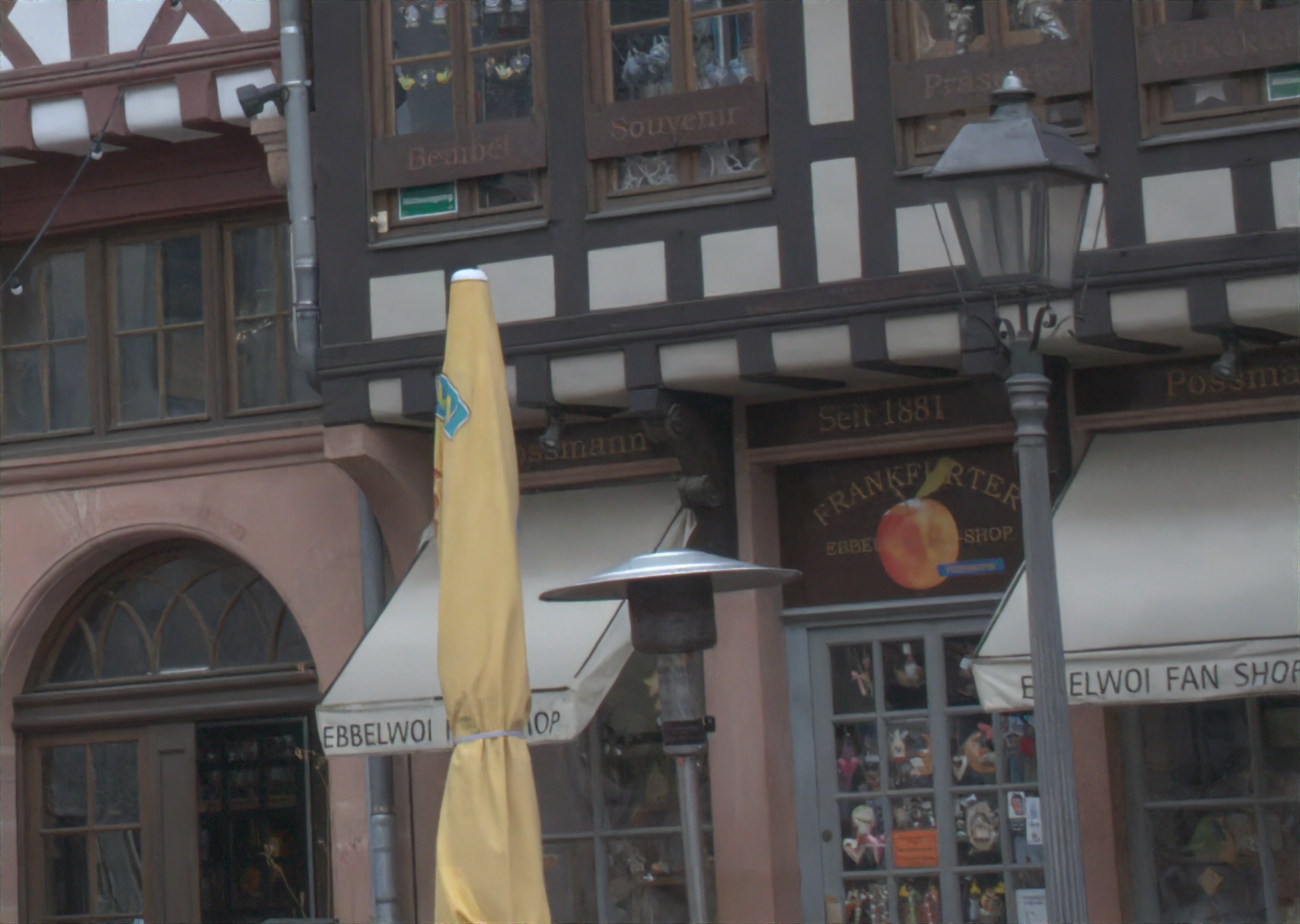}}};

\end{tikzpicture}
\end{figure*}

Examining an example image (image37) from the Darmstadt Noise Dataset (DND) under Poisson noise (Figure \ref{fig:DND 37_P}) shows that CMB3D performs visibly worse than Film-Unet, even on low levels of Poisson noise. Film-Unet retains more image detail (e.g. in the window reflections, signage details, texture in the columns).
Comparing to FFDNet (Figure \ref{fig:DND 37_P}), shows visibly similar performance at low noise levels, with Film-Unet retaining more detail (e.g. in the window details) as the noise increases. The metrics (Table \ref{tab:Comparison DND37}), show that Film-Unet outperforms the other methods at all noise levels and noise types.

\begin{figure*}[t]
\caption{Horses, image 40 CBSD68 dataset. Poisson Noise. $\sigma_{val}=0.05$ (left),  $\sigma_{val}=0.10$ (middle), $\sigma_{val}=0.20$ (right). Noisy image (bottom row), BM3D (2nd row), FFDNet (3rd row), Film-Unet (top row).}
\label{fig:Horses CBSD68 P}
\begin{tikzpicture}
\node at (0,0) {\includegraphics[width=6cm]{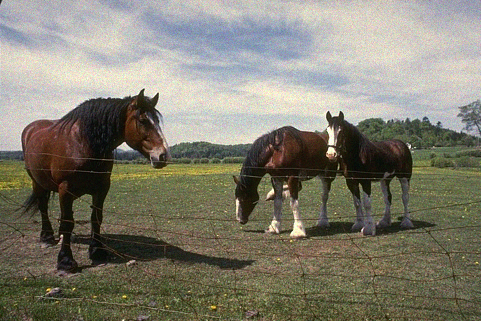}};
\node at (0,3.1) {\includegraphics[width=6cm]{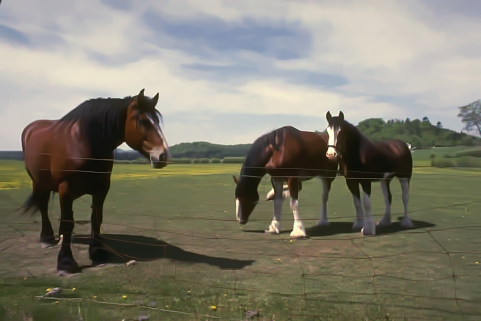}};
\node at (0,6.2) {\includegraphics[width=6cm]{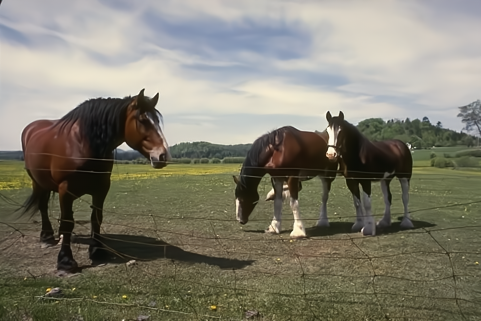}};
\node at (0,9.3) {\includegraphics[width=6cm]{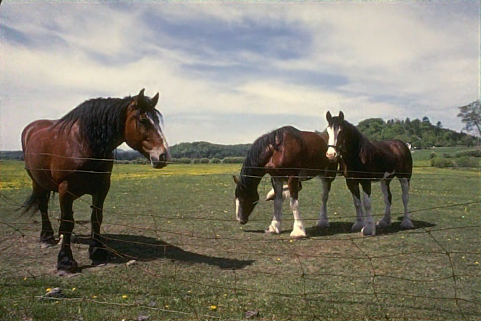}};

\node at (6,0) {\includegraphics[width=6cm]{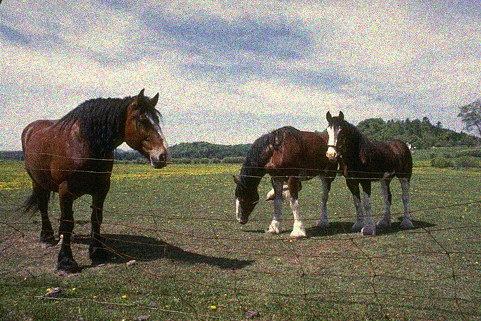}};
\node at (6,3.1) {\includegraphics[width=6cm]{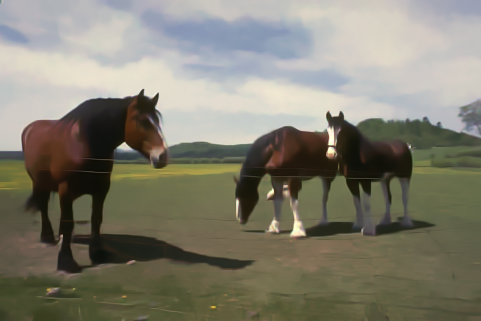}};
\node at (6,6.2) {\includegraphics[width=6cm]{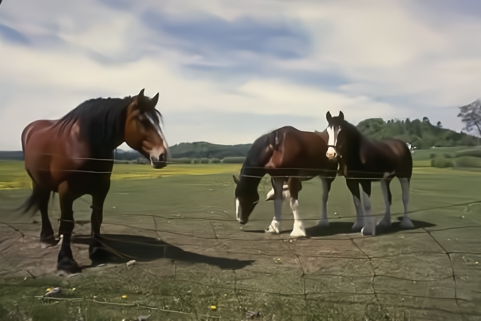}};
\node at (6,9.3) {\includegraphics[width=6cm]{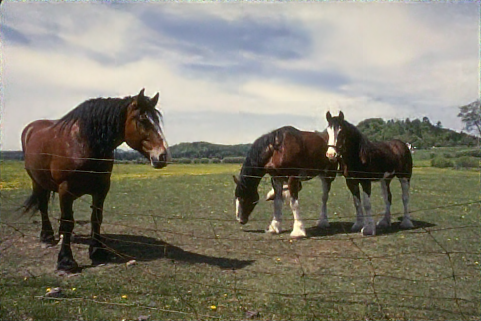}};

\node at (12,0) {\includegraphics[width=6cm]{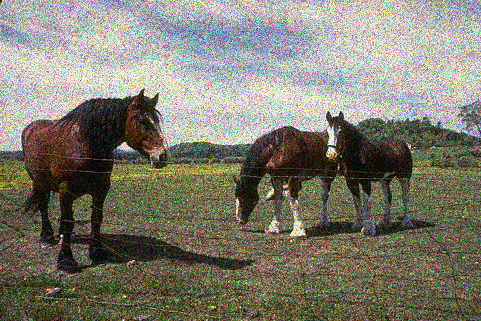}};
\node at (12,3.1) {\includegraphics[width=6cm]{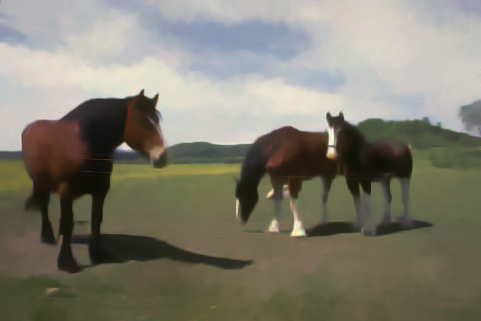}};
\node at (12,6.2) {\includegraphics[width=6cm]{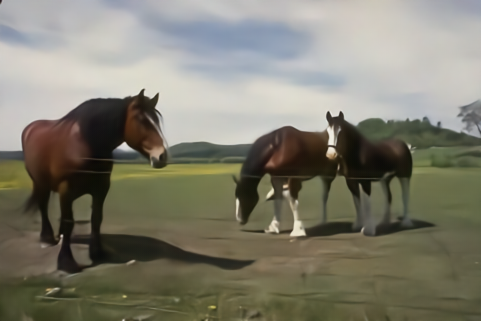}};
\node at (12,9.3) {\includegraphics[width=6cm]{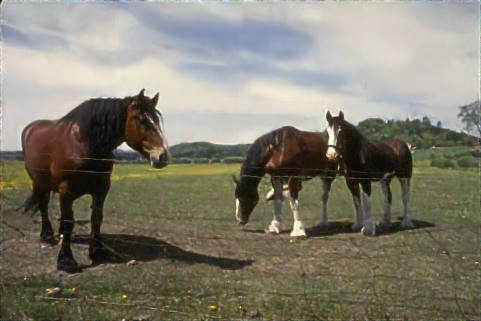}};
\end{tikzpicture}
\end{figure*}

Examining an example image (Horses, image 40) from the CBSD68 data set (Figure \ref{fig:Horses CBSD68 P}) shows that Film-Unet retains more image detail, even at low validation noise levels, whereas CBM3D and FFDNet are oversmooth, especially at high noise levels, where the foreground fence detail is no longer visible. Table \ref{tab:Comparison Horses} shows the corresponding metrics for this image, with Film-Unet outperforming CMB3D and FFDNet for higher levels of Poisson noise, and performing similarly to FFDNet at high Gaussian noise levels, but slightly worse than FFDNet under low levels of Gaussian noise.

\begin{table}[t]
\caption{Performance comparison for CBSD68 image 40, Horses}
\label{tab:Comparison Horses}
\begin{tabular}{|l|l|lll|}
\hline
$\sigma_{val}$ & Method & CBM3D & FFDNET & Film-Unet     \\ \hline
Gaussian &&&& \\ \hline
0.05 & PSNR &  30.2  &  33.6  &  32.0  \\ 
     & SSIM &  0.84  &  0.93  &  0.91  \\ \hline
0.1  & PSNR &  27.4  &  30.2  &  29.1  \\ 
     & SSIM &  0.72  &  0.86  &  0.82  \\ \hline
0.2  & PSNR &  25.4  &  26.6  &  26.8  \\ 
     & SSIM &  0.62  &  0.72  &  0.71  \\ \hline
Poisson &&&& \\ \hline
0.05 & PSNR &  29.8  &  34.9  &  34.5  \\ 
     & SSIM &  0.81  &  0.94  &  0.95  \\ \hline
0.1  & PSNR &  27.2  &  30.3  &  31.4  \\ 
     & SSIM &  0.69  &  0.83  &  0.90  \\ \hline
0.2  & PSNR &  25.5  &  26.9  &  28.6  \\ 
     & SSIM &  0.61  &  0.68  &  0.80  \\ \hline
\end{tabular}
\end{table}

\begin{table}[t]
\caption{Noise Performance comparison for DND image 37}
\label{tab:Comparison DND37}
\begin{tabular}{|l|l|lll|}
\hline
$\sigma_{val}$ & Method & CBM3D & FFDNet &Film-Unet     \\ \hline
Gaussian &&&& \\ \hline
0.05 & PSNR &  35.2  & 35.3  &  35.7  \\ 
     & SSIM &  0.91  & 0.92  &  0.92  \\ \hline
0.1  & PSNR &  33.1  & 34.1  &  34.2  \\ 
     & SSIM &  0.88  & 0.90  &  0.90  \\ \hline
0.2  & PSNR &  30.5  & 31.5  &  32.7  \\ 
     & SSIM &  0.84  & 0.87  &  0.88  \\ \hline
Poisson &&&&  \\ \hline
0.05 & PSNR &  35.2  & 35.1  &  37.7  \\ 
     & SSIM &  0.91  & 0.91  &  0.95  \\ \hline
0.1  & PSNR &  33.7  & 34.2  &  35.6  \\ 
     & SSIM &  0.88  & 0.90  &  0.92  \\ \hline
0.2  & PSNR &  31.7  & 33.0  &  34.0  \\ 
     & SSIM &  0.86  & 0.88  &  0.90  \\ \hline

\end{tabular}
\end{table}

\subsection{Analysis}
A comparison of PSNR and SSIM metric on 40 images from the S7 and CBSD68 datasets has shown that Film-Unet outperforms the state of the art comparison methods CBM3D and FFDNet, under both Gaussian noise and Poisson noise. The performance improvement under Poisson Noise is particularly noteworthy. An examination of some sample images shows that Film-Unet retains much more detail than CBM3D, especially at moderate to high noise levels. Compared to FFDNet, Film-Unet performs similarly under low levels of noise on some sample images, but significantly outperforms it at high levels of Poisson Noise and in general over multiple images.

Since Film-Unet is a neural network solution (as is FFDNet), its inference time performance is generally much faster than BM3D, especially for large images. By training conditionally on a range of noise types and levels, its inference time flexibility is improved compared to fixed noise model methods.  

In most experiments the network layers were jointly trained, however a two phase training of the Unet layers and Film layers separately shows that the Film layers can learn to generalise the fixed Unet denoiser to various noise levels by conditional training.

\section{Conclusion}
\label{sec:conclusion}
We have demonstrated a denoising model that is trained conditioned on a distribution of Poisson-Gaussian noise parameters such that the trained model can denoise images corrupted by Gaussian, Poisson or Poisson-Gaussian noise. It has been shown that matching the trained noise model to the noise corruption observed during validation results in better performance metrics PSNR, SSIM, preserving more authentic image detail. In particular, training on Poisson noise considerably improves the performance of Poisson input noise during validation. It has been shown that the Film layers can generalise a fixed Unet denoiser.

This may have particular relevance to applications such as medical imaging, in which Poisson-Gaussian noise corruption are prevalent where medical diagnosis may be aided by providing an improved and flexible denoiser with fast inference.

The influence of the noise parameters applied in the first phase of the two phase training experiment to the final, generalised, denoise performance is an interesting aspect of the work not dealt with here, that is worthy of future investigation.

\FloatBarrier 

\bibliographystyle{IEEEbib}
\bibliography{refs}

\end{document}